\documentclass[twocolumn,showpacs,superscriptaddress,article]{revtex4-1}
\usepackage[english]{babel}
\usepackage{amsmath,amssymb}
\usepackage{amsfonts}
\usepackage{mathtools}
\usepackage{siunitx}
\usepackage{libertine}
\usepackage[usenames,dvipsnames]{color}
\usepackage{hyperref}
\usepackage{blindtext}
\hypersetup{colorlinks}
\hypersetup{linkcolor=black, citecolor=black, urlcolor=blue}
\usepackage[utf8]{inputenc}
\usepackage{color}
\usepackage{subcaption}
\usepackage[normalem]{ulem}
\usepackage{xcolor}
\usepackage{graphicx}

\newcommand{\dif}{\mathrm{d}}

\usepackage{epstopdf}

\definecolor{cream}{RGB}{222,217,201}

\DeclareMathSizes{10}{9}{6}{4}

\begin{document}

\title{Coupled water, charge and salt transport in heterogeneous nano-fluidic systems}

\author{B. L. Werkhoven}
\affiliation{Institute for Theoretical Physics, Center for Extreme Matter and Emergent Phenomena, Utrecht University, Princetonplein 5, 3584 CC, Utrecht, The Netherlands}
\author{R. van Roij}
\affiliation{Institute for Theoretical Physics, Center for Extreme Matter and Emergent Phenomena, Utrecht University, Princetonplein 5, 3584 CC, Utrecht, The Netherlands}


\begin{abstract}
We theoretically study the electrokinetic transport properties of nano-fluidic devices under the influence of a pressure, voltage or salinity gradient. On a microscopic level the behaviour of the device is quantified by the Onsager matrix ${\bf L}$, a generalised conductivity matrix relating the local driving forces and the induced volume, charge and salt flux. Extending ${\bf L}$ from a local to a global linear-response relation is trivial for homogeneous electrokinetic systems, but in this manuscript we derive a generalised conductivity matrix ${\bf G}$ from ${\bf L}$ that applies also to heterogeneous electrokinetic systems. This extension is especially important in the case of an imposed salinity gradient, which gives necessarily rise to heterogeneous devices. Within this formalism we can also incorporate a heterogeneous surface charge due to, for instance, a charge regulating boundary condition, which we show to have a significant impact on the resulting fluxes. The predictions of the Poisson-Nernst-Planck-Stokes theory show good agreement with exact solutions of the governing equations determined using the Finite Element Method under a wide variety of parameters. Having established the validity of the theory, it provides an accessible method to analyse electrokinetic systems in general without the need of extensive numerical methods. As an example, we analyse a Reverse Electrodialysis "blue energy" system, and analyse how the many parameters that characterise such a system affect the generated electrical power and efficiency.   
\end{abstract}

\maketitle
\section{Introduction}

Over the past decades, the interest in nano- and microfluidics devices has significantly increased as these systems are able to control the transport of fluid, and thus dissolved solutes, with microscopic precision. The small scale of nanofluidic devices leads to novel properties compared to macrofluidic devices, allowing applications to a wide variety of different research fields 
\cite{Sparreboom2009,Schoh2008}. The great potential of such devices is additionally attested by  biological systems, which show an amazing control over permeability and selectivity of nanochannels 
\cite{Hille1978,Alcara2009,Martin2013,Schoh2008}.

\begin{figure}[!ht]
\centering
\includegraphics[width=0.45\textwidth]{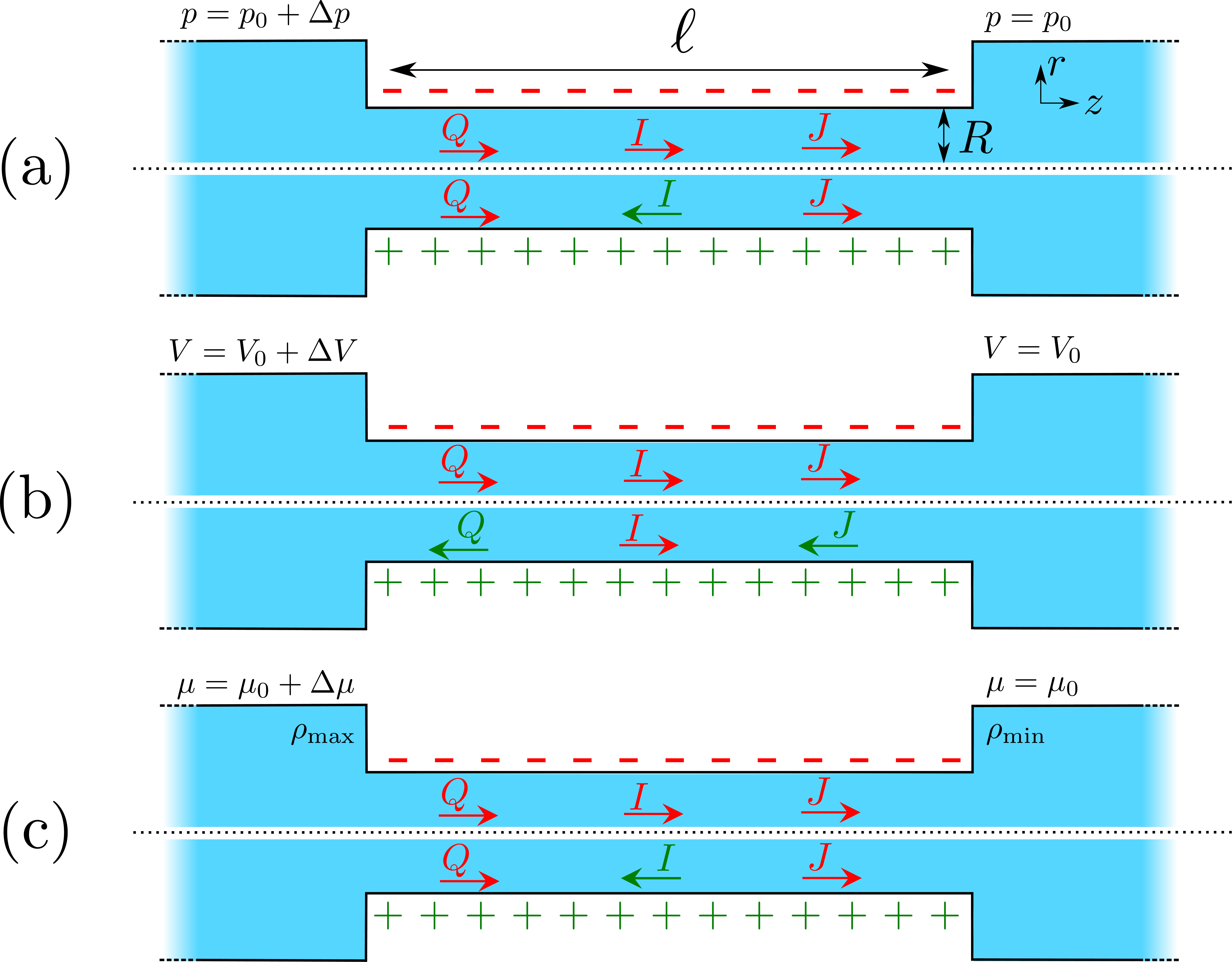}
\caption{A representation of a typical electrokinetic system with an imposed (a) pressure drop $\Delta p>0$, (b) electrostatic potential drop $\Delta V>0$ or (c) a chemical potential drop $\Delta \mu>0$ across a cylindrical channel with length $\ell$ and radius $R$. Here we consider both a positive (green) and negative (red) surface charge. The direction of the volumetric flow rate $Q$, electric current $I$ and solute flux $J$ depends on the sign of the surface charge and is indicated by the arrow and the colour. A red colour indicates that the flux is in the opposite direction to gradient of the applied driving force.}\label{fig:EKSystem}
\end{figure}

The unique properties of nano-fluidic devices derive ultimately from the relatively large surface to volume ratio. These properties make the field of nanofluidics of great importance for transport in porous materials such as porous rocks \cite{geology} and membranes \cite{membranehse}. Additionally, nano-fluidic devices offer new promising roads to desalination \cite{Elimelech2011}, DNA translocation 
\cite{Branton,He2016,Hatlo2011} and renewable energy harvesting \cite{Pennathur2007,Sales2010}. For instance, they have been used to convert hydrostatic energy into electric power  \cite{vanderHeyden2006,Lu2006} and to harvest energy from mixing salt and fresh water by Reverse Electrodialysis (RED) \cite{Post2008,Kim2010}, Pressure Retarded Osmosis (PRO)  \cite{Achilli2009,She2012,Straub2016} or Capacitive Double Layer Expansion (CDLE) \cite{Brogioli2009}. All of these nanofluidic devices are based on essentially the same system, composed of a channel with charged walls connecting two reservoirs with different reservoir conditions. Recent advances highlight the great potential for nanofluidics of Carbon Nanotubes (CNT) \cite{Park2014}, Boron Nitride Nanotubes (BNNT) \cite{Siria2013} and MoS$_2$ nanopores \cite{Feng2016}, which exhibit unique properties due to their small size and favourable electric properties. 

\section{Transport in electrokinetic systems}


Fig. \ref{fig:EKSystem} shows a representation of a typical electrokinetic system we will consider in this article: a cylindrical channel with a charged surface of length $\ell$ and radius $R$ connecting two bulk reservoirs containing a 1:1 electrolyte at room temperature. In this article we consider three different driving forces for transport, a pressure drop $\Delta p$, a voltage drop $\Delta V$ (electro-osmosis) and a salt chemical potential drop $\Delta \mu$ (i.e. a salt concentration drop $\Delta\rho$, diffusio-osmosis) over the channel. These driving forces can induce three different fluxes, i.e. currents integrated over a cross section: a volume or water flux $Q$ (m$^3$/s), more commonly known as the volumetric flow rate, a charge flux or electric current $I$ (A) and a net salt flux $J$ (s$^{-1}$).

Within linear response, we quantify the relation between the driving forces, $\Delta p$, $\Delta V$ and $\Delta\mu$, and generated fluxes, $Q$, $I$ and $J$, by a conductivity matrix ${\bf G}$,
\begin{equation}\label{eq:Ons2}
\begin{pmatrix} Q \\ I \\ J \end{pmatrix} = \frac{A}{\ell} {\bf G}
  \begin{pmatrix} \Delta p \\ \Delta V \\ \Delta\mu \end{pmatrix},
\end{equation}
with $A=\pi R^2$ the cross section area. The unique properties of nano-fluidic devices ultimately derive from the non-zero off-diagonal terms of ${\bf G}$, which highlight the highly interactive nature of nano-fluidic devices. If ${\bf G}$ is known, we can use Eq. \eqref{eq:Ons2} to calculate the fluxes generated by any set of imposed driving forces. For instance, an electric short-circuit or closed-circuit channel is obtained by electrically connecting the ends of the channel, such that $\Delta V=0$. If the salinities of the two reservoirs are different, i.e. diffusio-osmosis, Eq. \eqref{eq:Ons2} then gives the generated diffusio-osmotic electric current $I_{\rm DO}$ as
\begin{equation}\label{eq:IDO}
I_{\rm DO}=\dfrac{A}{\ell}\left(G_{21}\Delta p+G_{22}\Delta V+G_{23}\Delta\mu\right)=\dfrac{A}{\ell}G_{23}\Delta\mu,
\end{equation}
where we furthermore assumed a 'mechanical closed-circuit' condition, where water is free to flow (i.e. $\Delta p=0$). 

Alternatively, it is also possible to impose the flux instead of the applied potentials. For example, in an electric open-circuit channel the two reservoirs are not electrically connected and therefore no electric current can flow in steady state. In this case the flux $I=0$ is imposed instead of the potential drop, but then too Eq. \eqref{eq:Ons2} can be used. Since $\Delta \mu$ directly generates the current $I_{\rm DO}$ given by  Eq. \eqref{eq:IDO}, the only way to obtain a vanishing $I$ is for the system to develop a potential drop over the channel, commonly referred to as the diffusion potential $\Delta V_{\rm dif}$ 
\cite{Schoh2008}, such that the induced electro-osmotic current $I_{\rm EO}=\frac{A}{\ell}G_{22}\Delta V$ exactly cancels the diffusio-osmotic current $I_{\rm DO}$. The total current is simply the sum of the separate contributions, $I_{\rm total}=I_{\rm DO}(\Delta\mu)+I_{\rm EO}(\Delta V_{\rm dif})=0$, and we find
\begin{equation}\label{eq:Vdif}
\Delta V_{\rm dif}=-\dfrac{G_{23}}{G_{22}}\Delta\mu,
\end{equation}
The above two examples show that whether a flux or a driving force is imposed, in either case Eq. \eqref{eq:Ons2} can be used to calculate the remaining fluxes/driving forces. There is a great variety of imposed fluxes or driving forces that result in many different electrokinetic systems. Many of such electrokinetic systems are known by specific names, see Table \ref{tab:systems}, and Eq. \eqref{eq:Ons2} can be used for all possible combinations of driving forces.

\begin{table}[!ht]
\renewcommand{\arraystretch}{1.5}
\begin{tabular}{| p{3.45cm} || p{4.25cm} |}
\hline
Boundary Conditions & System \\
\hline \hline
$\Delta\mu= 0$, $\Delta p=0, \Delta V\neq0$ & Electro-osmosis\\
\hline
$\Delta\mu= 0$, $\Delta p\neq0, I=0$ & Streaming potential\\
\hline
$\Delta\mu\neq 0$, $\Delta p=0, I=0$ & Membranes/diffusio-osmosis \\
\hline
$\Delta\mu\neq 0$, $\Delta p\neq 0, I=0,$& Pressure Retarded Osmosis \& desalination\\
\hline
$\Delta \mu =0$, $\Delta p\neq 0$, $I\neq 0$ & Mechanical energy conversion\\
\hline
$\Delta \mu \neq0$, $\Delta p= 0$, $I\neq 0$ & Reverse Electrodyalisis\\
\hline
$\Delta \mu =0$, $Q=0$, $\Delta V\neq 0$ & Capacitive Double Layer Expansion\\
\hline
\end{tabular}
\caption{Collection of electro-kinetic systems and the associated boundary conditions, with $\Delta p$, $\Delta V$, and $\Delta\mu$ the pressure, voltage and chemical potential drop across the channel, and $I$ and $Q$ the electric current and volumetric flow rate through the channel.}\label{tab:systems}
\end{table}

In this article, we will show how we can obtain the conductivity matrix ${\bf G}$ from a well-known microscopic linear response theory based by the Onsager matrix ${\bf L}$, which we will calculate analytically within the Poisson-Boltzmann formalism. We then show how to extend ${\bf L}$, which is in essence a local linear-response equation, to ${\bf G}$, which is a global linear-response equation.
In order to validate our method, we compare predictions of  Eq. \eqref{eq:Ons2} with solutions of the Poisson-Nernst-Planck-Stokes equations obtained using Finite Element Method (FEM). While FEM results are typically more precise, the great advantage of the proposed method is that these are much easier to implement and do not require complicated numerical techniques, and can thus be more easily used to analyse more complex nanofluidic systems. As an example, we will use the generalised conductivity matrix ${\bf G}$ to show how to incorporate a charge regulation mechanism with a salinity gradient, and compare predictions of the generated current with experiments on Boron Nitride Nanotubes \cite{Siria2013}. The proposed framework provides a general formalism to investigate all electrokinetic systems as listed in Table \ref{tab:systems}, but as an example we will focus on  ${\bf G}$ to analyse an electrokinetic system using Reverse Electrodyalisis under a wide variety of parameters without the need for extensive numerical calculations with FEM. This analysis highlights the convenience and utility of the conductivity matrix ${\bf G}$ for nanofluidics and electrokinetic systems in general.

\section{Linear response electrokinetic}

A well-known method to describe the transport properties of nano-fluidic channels is by the so-called Onsager matrix ${\bf L}$ \cite{Onsager1, Onsager2, Fair1971, Peters2016, Rastogi1993}, which relates the local driving forces to the generated fluxes. Within linear response theory, the induced fluxes are linear in the driving forces
\begin{equation}\label{eq:Ons}
\begin{pmatrix} Q \\ I \\ J-2\rho_sQ \end{pmatrix} = A{\bf L}
  \begin{pmatrix} -\partial_z p \\ -\partial_z V \\ -\partial_z\mu \end{pmatrix},
\end{equation}
where $\partial_z$ is the derivative with respect to the lateral Cartesian coordinate $z$ and ${\bf L}$ is a symmetric 3$\times$3 matrix. For electrokinetic systems, composed of channels with charged walls in contact with an electrolyte, ${\bf L}$ can be determined fully analytically with the Poisson-Boltzmann formalism (see Supplementary Information). The flux associated to $\partial_z\mu$ is the excess salt flux $J_{\rm exc}=J-2\rho_sQ$, the total salt flux $J$ minus the bulk advective salt flux, with $\rho_s$ the salt concentration (salinity) at the channel axis. Defining the Onsager matrix in terms of $J_{\rm exc}$ rather than $J$ ensures that ${\bf L}$ is symmetric (see Supplementary Information for more information) 
\cite{Onsager1,Onsager2,Fair1971}. 

The disadvantage of  Eq. \eqref{eq:Ons}, however, is that it relates the \emph{local} driving forces to the fluxes, while Eq. \eqref{eq:Ons2} relates the \emph{global} driving forces to the fluxes. Since the global rather than the local driving forces are experimentally imposed or measured, in order for  Eq. \eqref{eq:Ons} to be useful it must be extended to the same form as  Eq. \eqref{eq:Ons2}. This is straightforward if ${\bf L}$ is constant throughout the channel, since then we can simply integrate  Eq. \eqref{eq:Ons} along the length of the channel and find that ${\bf L}={\bf G}$. This is the case when a non-zero $\Delta p$ and $\Delta V$ is imposed, since only under extreme circumstances do these influence the properties of the channel. However, since the properties of the electric double layer are strongly affected by the salinity $\rho_s$, a non-zero $\Delta \mu$ necessarily leads to a laterally varying salinity $\rho_s$ and thus a laterally varying ${\bf L}$.  In that case, therefore, it is no longer clear how to convert  Eq. \eqref{eq:Ons} to a global equation, except in the case of a small relative change in salinity across the channel. If, however, the salinity changes for example from 20 mM to 500 mM, as is the case for fresh to sea water, a clear method is required to obtain the fluxes from ${\bf L}$. 

\subsection{Global linear response}

One method to obtain the fluxes as a function of the global driving forces as in Eq. \eqref{eq:Ons2}, is to resolve  Eq. \eqref{eq:Ons} for every location $z$ for a given value of the flux. Such adjustments have been successfully incorporated before \cite{Fair1971, Peters2016}, but since the local driving forces are in principle unknown, this method gives the driving force as a function of the flux instead of the global driving forces as  Eq. \eqref{eq:Ons2}. Since the latter is clearly preferable, this method becomes rather cumbersome. Here we show how to extend ${\bf L}$ to ${\bf G}$, while retaining the convenience of  Eq. \eqref{eq:Ons2}.

\begin{figure}[!ht]
\centering
\includegraphics[width=0.5\textwidth]{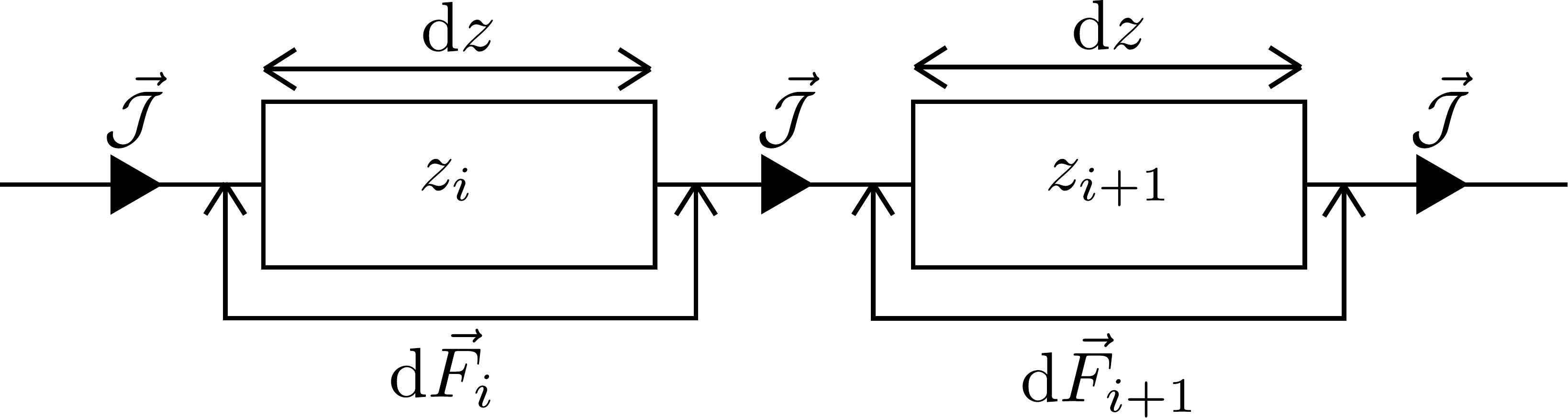}
\caption{Schematic representation of an electrokinetic system divided in infinitessimally small segments of width d$z$, with an applied driving force d$\vec{F}_i$ over each segment and a flux $\vec{\mathcal{J}}$ through each segment. Each segment ${\bf L}(z_i)$ and d$\vec{F}_i$ can locally take different values, but $\vec{\mathcal{J}}$ is a spatial constant in steady state.}\label{fig:fluxes}
\end{figure}
In order to obtain ${\bf G}$ from a heterogeneous ${\bf L}(z)$ we start from the condition that all fluxes $Q$, $I$ and $J$ are, in steady state and for non-leaky channels, constant throughout the channel (independent of $z$). In order to calculate the fluxes as a function of the global driving forces, we divide the system into infinitesimally small segments of width d$z$, schematically represented in Fig. \ref{fig:fluxes}, and apply the Onsager equation, Eq. \eqref{eq:Ons}, for each segment
\begin{equation}
\vec{\mathcal{J}}=A\left({\bf L}(z_i)+{\bf L}^{\rm adv}(z_i)\right)\cdot\left(-\dfrac{{\rm d}\vec{F}_i}{{\rm d}z}\right),
\end{equation}
where $\vec{\mathcal{J}}=(Q, I, J)$ and d$\vec{F}_i$/d$z$=($\partial_z p$, $\partial_z V$, $\partial_z \mu)\big|_{z=z_i}$ is a vector that contains all fluxes and driving forces over the $i$th segment, respectively. Furthermore, ${\bf L}^{\rm adv}$ is the bulk advective salt flux, which accounts for the difference between $J$ and $J_{\rm exc}$, 
\begin{equation}
{\bf L}^{\rm adv}(z)=2\rho_s(z)\begin{pmatrix} 0 & 0 & 0 \\ 0 & 0 & 0 \\ L_{11} & L_{12} & L_{13} \end{pmatrix},
\end{equation}
with $\rho_s(z)$ the salinity at the channel axis ($r=0$) at lateral position $z$. Note that ${\bf L}^{\rm adv}$ simply adds the local advective salt flux $2\rho_sQ$ to the excess salt flux, since $J=J_{\rm exc}+2\rho_sQ$. This contribution must be included because in steady state, by virtue of the incompressibility of water and due to charge and ion number conservation, $Q$, $I$ and $J$ and thus $\vec{\mathcal{J}}$ can not depend on $z$ ($J_{\rm exc}$ can in principle depend on $z$). We can obtain the global driving forces by summing (integrating in the continuum limit) all d$\vec{F}_i$,
\begin{equation}
\Delta\vec{F}=-\int\limits_0^{\ell}{\rm d}z\dfrac{{\rm d}\vec{F}}{{\rm d}z}=\dfrac{1}{A}\int\limits_0^{\ell}{\rm d}z\left({\bf L}+{\bf L}^{\rm adv}\right)^{-1}\cdot \vec{\mathcal{J}},
\end{equation}
where $\Delta\vec{F}=(\Delta p,\Delta V, \Delta\mu)$ is the vector containing all global driving forces. Inverting this equation we obtain the (constant) fluxes $\vec{\mathcal{J}}$ as a function of the global driving forces $\Delta\vec{F}$,
\begin{equation}
\vec{\mathcal{J}}=A\left(\int\limits_0^{\ell}{\rm d}z\left({\bf L}+{\bf L}^{\rm adv}\right)^{-1}\right)^{-1}\cdot\Delta\vec{F}\equiv\dfrac{A}{\ell}{\bf G}\cdot\Delta\vec{F}.
\end{equation}
Here, the conductivity matrix ${\bf G}$, as defined in  Eq. \eqref{eq:Ons2}, can thus be obtained from ${\bf L}$ as
\begin{equation}\label{eq:cond}
{\bf G}^{-1}=\dfrac{1}{\ell}\int\limits_0^{\ell}\dif z\big[({\bf L}(\rho_s(z))+{\bf L}^{\rm adv}(\rho_s(z))\big]^{-1}.
\end{equation}
As stated before, the Onsager matrix ${\bf L}$ can be determined analytically within Poisson-Boltzmann theory, and we can subsequently use Eq. \eqref{eq:cond} to find the conductivity matrix ${\bf G}$.

\begin{figure}[!ht]
\centering
\includegraphics[width=0.4\textwidth]{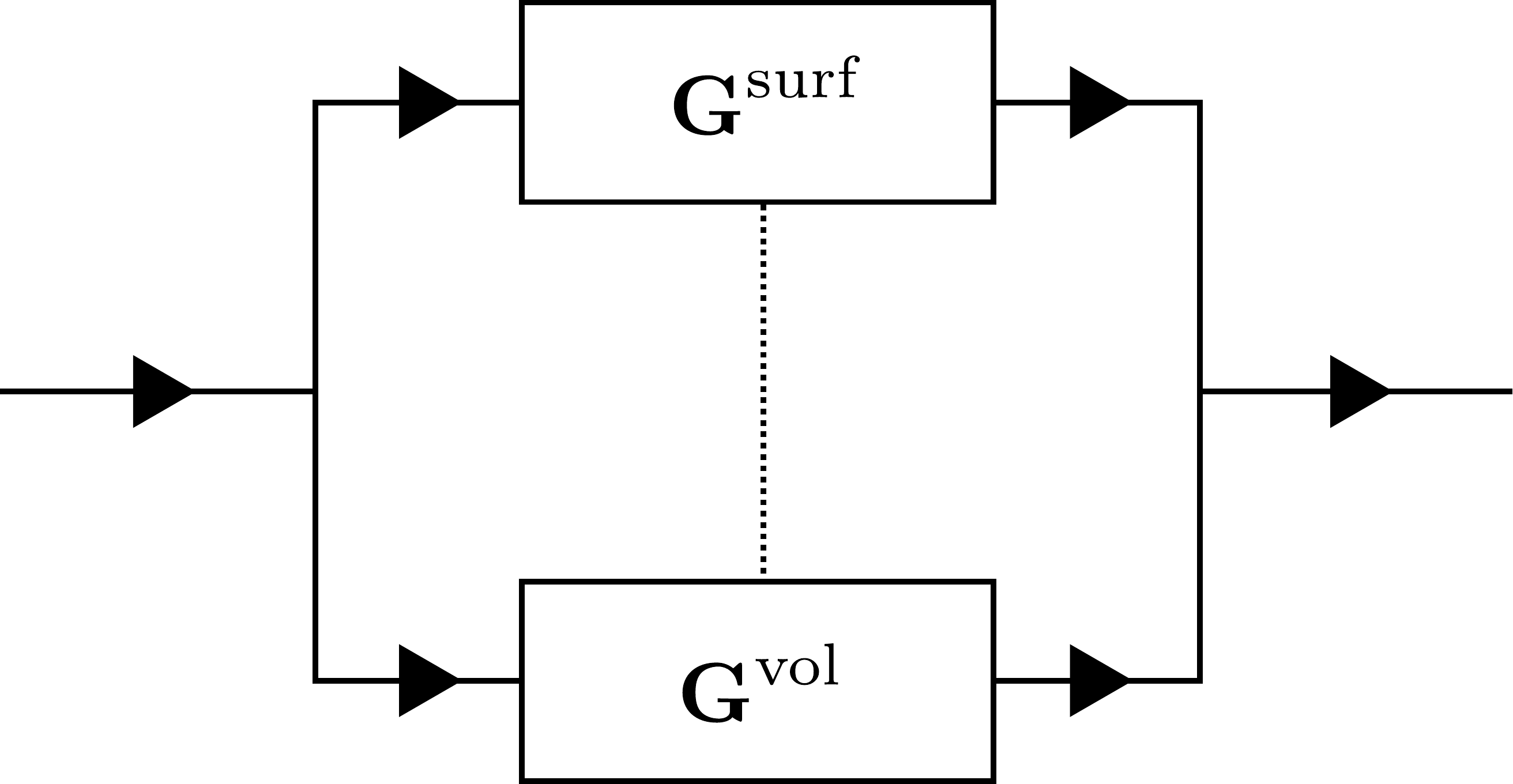}
\caption{Analogue electrical circuit representation of an electrokinetic system.}\label{fig:para}
\end{figure}

However, we can significantly simplify Eq. \eqref{eq:cond} by splitting the contributions to ${\bf L}$ in a volume (${\bf L}^{\rm vol}$) and a surface (${\bf L}^{\rm surf}$) contribution, 
\begin{equation}
{\bf L}={\bf L}^{\rm vol}+{\bf L}^{\rm surf},
\end{equation}
where ${\bf L}^{\rm vol}$ consists of all contributions of the order $R^0$ (or higher) and ${\bf L}^{\rm surf}$ consists of all terms proportional $R^{-1}$, with $R$ the channel radius. We then treat the volume and surface contributions as separate conductors incorporated in a parallel circuit. To illustrate this, we consider an analogous electrical circuit where two resistors (conductors) are connected in parallel, as in Fig. \ref{fig:para}.
In principle, the induced fluxes $Q$, $I$ and $J$ can flow via the EDL, represented by ${\bf G}^{\rm surf}$ or via the region outside the EDL, represented by ${\bf G}^{\rm vol}$ (each a sequence of many infinitesimally small conductors as in Fig. \ref{fig:fluxes}). These two are, in general, connected, represented by the dashed line (to be precise, every infinitesimal conductor is connected to its volume/surface counterpart). We can, however, significantly simplify the system by disconnecting the surface and volume fluxes (i.e. removing the dashed line in Fig. \ref{fig:para}), which can intuitively be understood by realising that all radial components of the fluxes are small or negligible (such that the interchange between volume and surface is also small). We expect this simplification to break down for small aspect ratios $\ell/R$ and/or large heterogeneities across the channel.


The advantage of separating the volume and surface contributions is that the total conductance is now determined by the sum of the two separate conductances (note that ${\bf G}^{\rm vol}$ and ${\bf G}^{\rm surf}$ themselves can still originate from a laterally heterogeneous ${\bf L}^{\rm vol}$ and ${\bf L}^{\rm surf}$ respectively). We can analytically calculate ${\bf G}^{\rm vol}$, by evaluating Eq. \eqref{eq:cond} with ${\bf L}^{\rm surf}={\bf 0}$ (see Supplementary Information for a derivation). On the other hand, it is not possible to determine ${\bf G}^{\rm surf}$ analytically in the same way as ${\bf G}^{\rm vol}$. In order to obtain an analytic expression we approximate ${\bf G}^{\rm surf}$ by ${\bf L}^{\rm surf}$ evaluated at the average salinity $\bar{\rho}=\frac{1}{2}(\rho_{\rm min}+\rho_{\rm max})$,
\begin{equation}
{\bf G}^{\rm surf}=\left(\dfrac{1}{\ell}\int\limits_0^{\ell}\dif z{\bf L}^{\rm surf}(z)^{-1}\right)^{-1}\approx {\bf L}^{\rm surf}(\bar{\rho}),
\end{equation}
where $\rho_{\rm min}$ and $\rho_{\rm max}$ are the salt concentration of the low and high salinity reservoir respectively. Note that we could also have chosen the geometric mean $\bar{\rho}_{\rm geom}= \sqrt{\rho_{\rm min}\rho_{\rm max}}$, but we found the arithmetic mean to provide (slightly) more accurate predictions compared to the FEM results. The total conductivity matrix ${\bf G}$ can then be approximated as
\begin{equation}\label{eq:Gana}
{\bf G}\approx {\bf G}^{\rm vol}+ {\bf G}^{\rm surf}\approx {\bf G}^{\rm vol}+{\bf L}^{\rm surf}(\rho_s=\bar{\rho}),
\end{equation}
with ${\bf G}^{\rm vol}$ given in Eq. \eqref{eq:Gvol} in Supplementary Information. As we will see below, Eq. \eqref{eq:cond} can accurately predict the FEM results over a large range of parameter values, and Eq. \eqref{eq:Gana} is surprisingly accurate given the simplifications involved.

One significant advantage of the above formalism is that it is straightforward to also incorporate lateral heterogeneities other than a salinity gradient. For example, we will consider BNNTs and CNTs in this article, which obtain their surface charge from the adsorption of an OH$^-$ ion. Because OH$^-$ carries a net charge, the amount of OH$^-$ adsorption depends on the surface charge itself via a mechanism known as charge regulation 
\cite{chargeregulation, Chan1983,hunter}, and can be expressed as a Langmuir-type relation
\begin{equation}\label{eq:CR}
\sigma(z)=z_{\sigma}\Gamma\left(1+10^{-{\rm pH+pK}}e^{-e\psi_0(z)/k_{\rm B}T}\right)^{-1},
\end{equation}
where $z_{\sigma}$ is the valency of the surface charge ($z_{\sigma}=-1$ for OH$^-$ adsorption), pK the reaction constant of the charging mechanism, $\Gamma$ is the areal density of chargeable surface sites, and $\psi_0$ the surface potential. The relation between $\sigma$ and $\psi_0$ depends on the (local) salinity, given by the Poisson-Boltzmann formalism (see Supplementary Information), such that Eq. \eqref{eq:CR} is a self-consistency relation for the local surface charge $\sigma(z)$. Note that, for simplicity, we leave out a Stern layer capacitance from Eq. \eqref{eq:CR}. Since $\psi_0$ is a function of $\rho_s$, Eq. \eqref{eq:CR} implies a heterogeneous surface charge in the case of $\Delta\mu\neq 0$ (diffusio-osmosis), which is straightforwardly included in the above formalism. The charge-regulation boundary condition, however, can significantly affect the resulting fluxes, as we will shown below, and has been shown to be important for the interpretation of measurements on CNTs 
\cite{Biesheuvel2016, Secchi2016}.

\subsection{Entrance Effects}

One final point to address concerning ${\bf G}$ is that a density profile $\rho_s(z)$ is required in order to use Eq. \eqref{eq:cond}. A straightforward example is of course a purely diffusive (i.e. linear) profile, although one should keep in mind that this is not necessarily accurate because the profile can be influenced by an advective fluid flow or an electric field 
\cite{Lee2014}. 
\begin{figure}[!ht]
\centering
\includegraphics[width=0.48\textwidth]{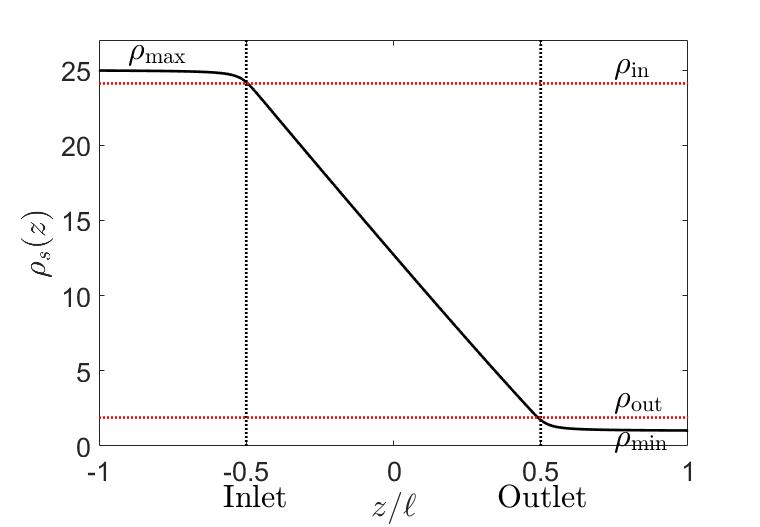}
\caption{Density profile at the axis of the channel calculated with FEM (black full line) for $R=60$ nm and $\ell=1500$ nm. The dashed red lines indicate the inlet and outlet salinity $\rho_{\rm in}\approx 2$ mM and $\rho_{\rm out}\approx 24$ mM, and the black dashed lines indicate the location of the inlet $\left(z=-\frac{1}{2}\ell\right)$ and outlet $\left(z=-\frac{1}{2}\ell\right)$.}\label{fig:sal}
\end{figure}
The density profile in a finite channel is, however, also affected by entrance effects. Due to the finite size of the channel, the salinity at the in- and outlet of the channel is not exactly equal to reservoir salinities $\rho_{\rm max}$ and $\rho_{\rm min}$. However, the salinity gradients in the far field of the reservoirs vanish, resulting in a region at the in- and outlet, outside the channel, with a salinity different from $\rho_{\rm max}$ and $\rho_{\rm min}$. This is confirmed by FEM calculations, which show that the salinity at the inlet is lower than $\rho_{\rm max}$, and the salinity at the outlet is higher than $\rho_{\rm min}$ (see Fig. \ref{fig:sal}). The corrections are not large, but one must keep in mind that the conductivity of the channel is, according to  Eq. \eqref{eq:cond}, most strongly affected by the smallest conductivity, i.e. the low salinity side. A small correction at the outlet can thus have significant effects on the total conductivity. 

Fig. \ref{fig:sal} shows the salinity at the channel axis as determined from FEM solutions of the PNPS equations. Even for a very needle-shaped channel ($\ell/R=25$), the in- and outlet salinities clearly differ from the reservoir salinities. The effect becomes more pronounced for shorter and/or wider channels with a small aspect ratio. For example, for $\ell/R=5$ the outlet salinity is a factor 4 larger than $\rho_{\rm min}$ (see Supplementary Information).  We denote the inlet salinity as $\rho_{\rm in}$ and the outlet salinity as $\rho_{\rm out}$, which now explicitly depend on $R$ and $\ell$ due to the entrance effects (see Supplementary Information for derivation). This correction is similar (although not equal) to the so-called access resistance \cite{He2016}, as it also slightly adjusts the salinity gradient. The chemical potential drop over the channel $\Delta\mu_{\rm ch}$ is consequently not equal to the chemical potential difference $\Delta\mu=k_{\rm B}T\log\dfrac{\rho_{\rm max}}{\rho_{\rm min}}$ between the two reservoirs, but actually
\begin{equation}
\Delta\mu_{\rm ch}=k_{\rm B}T\log\dfrac{\rho_{\rm in}}{\rho_{\rm out}}.
\end{equation} 
The distinction between $\Delta\mu$ and $\Delta\mu_{\rm ch}$, $\rho_{\rm max}$ and $\rho_{\rm in}$ and $\rho_{\rm min}$ and $\rho_{\rm ouot}$ is a small but significant one, the more so for shorter and wider channels.


In this article we assume a linear profile 
\begin{equation}\label{eq:sal}
\rho_s(z)=\rho_{\rm in}-\left(\frac{1}{2}+\dfrac{z}{\ell}\right)(\rho_{\rm in}-\rho_{\rm out}),
\end{equation} 
from $\rho_{\rm in}$ to $\rho_{\rm out}$, for $-\frac{1}{2}\ell<z<\frac{1}{2}\ell$, where the in- and outlet salinities are given by (see Supplementary Information for derivation)
\begin{equation}
\rho_{\rm out}\approx\rho_{\rm min}+\frac{R}{\ell+2R}\Delta \rho, \hspace{4mm} \rho_{\rm in}\approx\rho_{\rm max}-\frac{R}{\ell+2R}\Delta \rho,
\end{equation} 
with $\Delta\rho=\rho_{\rm max}-\rho_{\rm min}$ the salinity difference between the reservoirs. Note that  Eq. \eqref{eq:sal} introduces an explicit dependence on the channel length $\ell$ in the formalism via $\rho_{\rm in}$ and $\rho_{\rm out}$, as has indeed been shown to be a non-trivial parameter for diffusio-osmosis \cite{Cao2017}. Only for infinitely long channels do we find that $\rho_{\rm in}=\rho_{\rm max}$ and $\rho_{\rm out}=\rho_{\rm min}$. In general, a salinity profile will be affected by the fluid flow and can be found by solving the convection-diffusion equation. However, the resulting exponential profile reduces to a linear profile if the fluid flow is not too large, more precisely if the Peclet number Pe=$\frac{Q\ell}{\pi R^2 D}=(G_{11}\Delta p+G_{12}\Delta V+G_{13}\Delta\mu_{\rm ch})/D$ is significantly smaller than unity. This is typically the case for diffusio-osmosis, except for very large slip lengths (exceeding tens of nanometers). In that case, the salinity profile must be adjusted to a profile predicted by a diffusion-convection system. 

\subsection{The Onsager matrix}

\begin{figure*}[!ht]
\begin{equation}\label{eq:Onscoef}
\begin{aligned}
L_{11}^{\rm vol}&=-\frac{R^2}{8\eta}\left(1+\frac{4b}{R}\right) \qquad &L_{11}^{\rm surf}&=0\\
L_{12}^{\rm vol}&=-\frac{\epsilon\psi_0+be\sigma}{\eta} \qquad &L_{12}^{\rm surf}&=z_{\sigma}\frac{e\lambda_D}{2\pi\lambda_B \eta R}P_1,\\
L_{13}^{\rm vol}&=\frac{1}{4\pi\lambda_B\eta}\left(\frac{b}{\lambda_D}P_2+P_3\right) \qquad&L_{13}^{\rm surf}&=-\frac{\lambda_D}{8\pi\lambda_B\eta R}P_4,\\
L_{22}^{\rm vol}&=\frac{2 De^2}{k_{\rm B}T}\rho_s \qquad &L_{22}^{\rm surf}&=\dfrac{2De^2}{k_{\rm B}T}   \left(\frac{2\rho_s\lambda_D}{R}P_2\left(1+\frac{k_{\rm B}T}{2\pi\lambda_B\eta D}\right)-\beta\frac{\sigma}{R} \right)+2e^2\frac{b}{R}\frac{\sigma^2}{\eta},\\
L_{23}^{\rm vol}&=\beta\frac{2 De}{k_{\rm B}T}\rho_s \qquad &L_{23}^{\rm surf}&=-\dfrac{2De}{k_{\rm B}T}\left(\frac{\sigma}{R}-\beta\frac{2\rho_s\lambda_D}{R}P_2\right)-\frac{e}{2\pi\lambda_B\lambda_D\eta R}\left(\frac{z_{\sigma}}{4\pi\lambda_B}P_5+b\sigma P_2\right),\\
L_{33}^{\rm vol}&=\frac{2 D}{k_{\rm B}T}\rho_s \qquad &L_{33}^{\rm surf}&=\dfrac{2D}{k_{\rm B}T}\left(\frac{2\rho_s\lambda_D}{R}P_2-\beta\frac{\sigma}{R}\right)+\frac{\rho_s\lambda_D}{\pi\lambda_B\eta R}\left(2P_2-4P_3+\frac{b}{\lambda_D}P^2_2\right).
\end{aligned}
\end{equation}
\end{figure*}

So far we have explained how to extend the local linear response Onsager matrix ${\bf L}$ to a global linear response conductivity matrix ${\bf G}$. As mentioned, ${\bf L}$ originates partially from the surface charge of the channels walls, which can be either imposed or spontaneously originate from chemi- or physisorption of ions. This surface charge attracts oppositely charged ions to, and repels equally charged ions from, the surface, giving rise to a non-zero space charge close to the surface called the Electric Double Layer (EDL). The EDL consists of charge and concentration gradients perpendicular to the surface which extend into the fluid over a typical distance of the Debye length $\lambda_D$, and therefore affects the fluxes parallel to the surface. We assume here that the EDL is in its equilibrium configuration before the driving forces are applied, since the EDL equilibrates typically on a timescale of the order of nano- to microseconds \cite{Bazant2004}. This allows us to use the solutions of Poisson-Boltzmann formalism to derive ${\bf L}$.

In this article we will consider an electrokinetic system as depicted in Fig. \ref{fig:EKSystem}, with length $\ell$, radius $R$, salinity $\rho_s(z)$ given by Eq. \eqref{eq:sal} and surface charge $\sigma$. The fluid flow is determined by the Stokes equation with an electric body force and the incompressibility condition 
\cite{hunter},
\begin{equation}\label{eq:Stokesch3}
-\nabla p + \eta \nabla^2 {\bf u}+e(\rho_+-\rho_-){\bf E}=0, \qquad \nabla \cdot {\bf u}=0,
\end{equation}
with the slip boundary condition
\begin{equation}
b\partial_r u_z(r=R)=u_z(r=R),
\end{equation}
with the channel axis oriented in the $z$ direction. Here $p$ is the hydrostatic pressure (i.e. sum of the partial solvent pressure and osmotic pressure due to the ions), ${\bf u}$ the fluid velocity vector,  $\eta$ the viscosity, ${\bf E}$ the electric field, $e$ the proton charge, $\rho_{\pm}$ the local cation/anion number density, $b$ the slip length and $r\in [0,R]$ the coordinate normal to the surface. The ion fluxes are given by the Nernst-Planck equation 
\cite{hunter},
\begin{equation}\label{eq:PNP}
{\bf j}_i=-D_i\nabla\rho_i+z_i\frac{D_ie}{k_{\rm B}T}\rho_i{\bf E}+\rho_i{\bf u},
\end{equation} 
with $k_{\rm B}$ the Boltzmann constant, $T$ the temperature and $\rho_i$, $D_i$, $z_i$ the density, the diffusion constant and the valency of ion species $i=\pm$, respectively. We consider in this article a 1:1 salt, as this makes it possible to solve all equations analytically (although these are straightforwardly extended to a $z$:$z$ salt). We obtain the fluxes as
\begin{equation}
\begin{aligned}
Q&=2\pi\int\limits_0^R\dif r r u_z,\\
I&=2\pi e\int\limits_0^R\dif r r (j_{+,z}-j_{-,z}),\\
J&=2\pi\int\limits_0^R\dif r r (j_{+,z}+j_{-,z}),\\
\end{aligned}
\end{equation} 
for a cylindrical geometry. Note that $J$ is the total and not the excess salt flux $J_{\rm exc}$.  


By combining the above equations with the solutions of Poisson-Boltzmann formalism for a 1:1 salt  
\cite{Schoh2008,overbeek}, the full 3$\times$3 Onsager matrix can be determined analytically. The majority of the matrix elements of ${\bf L}$ are already known, although we do find a contribution to $L^{\rm surf}_{23}$, the non-advective contributions of Eq. \eqref{eq:Onscoef}, that appears to have been overlooked in previous studies \cite{Mouterde2018,Siria2013}. It is an important contribution that cannot be ignored, and is in fact required by the symmetry of ${\bf L}$. This term is intimately linked to the heterogeneity of the EDL: since the Debye length $\lambda_D$ is a function of $z$, diffusio-osmosis generates a lateral component to the electric field which contributes to the generated fluxes (see Supplementary Information for detailed discussion of this subtle contribution). For the sake of completeness, however, we present not just $L_{32}$ but the full 3$\times$3 matrix. 

Eq. \eqref{eq:Onscoef} shows the Onsager matrix elements, with $\lambda_B=\frac{e^2}{4\pi\epsilon k_{\rm B} T}$ the Bjerrum length and $\lambda_D=\left(8\pi\lambda_B\rho_s\right)^{-1}$ the Debye length, $\epsilon$ the permittivity of water, $\psi_0$ the surface potential, $z_{\sigma}$ the sign of the surface charge, $D=\frac{1}{2}(D_++D_-)$ the average ion diffusion constant and $\beta=\frac{D_+-D_-}{D_++D_-}$ the mobility mismatch. The constants $P_i$ are positive numbers and function of $\rho_s$, $\sigma$ and $\psi_0$ only. For small surface charge, $2\pi\lambda_B\lambda_D\sigma\ll 1$, all these constants scale as $P_i\sim\sigma^2\sim\phi_0^2$, while for large surface charge, $2\pi\lambda_B\lambda_D\sigma\gg 1$, $P_1\approx\pi^2/2$, $P_2\sim \sigma$, $P_3\sim|\phi_0|$, $P_4\approx\pi^2/4$ and $P_5\sim\sigma$. These constant are solutions to rather involved integrals, and the full expressions and their derivations can be found in the Supplementary Information. Note that $L^{\rm vol}_{12}$ and $L^{\rm vol}_{23}$ change sign if $\sigma$ changes sign, while $L^{\rm vol}_{13}$ does not. This is directly reflected in Fig. \ref{fig:EKSystem}, which shows that $Q$ and $J$ are always in the same direction while the direction of $I$ with respect to $Q$ and $J$ depends on the sign of $\sigma$. 

Most elements are known by specific names, for example in the context of electro-osmosis 
\cite{hunter} and diffusio-phoresis \cite{Prieve1984}; $L_{11}$ is inversely proportional to the fluidic impedance $Z_{\rm ch}=\frac{\ell}{\pi R^2 L_{11}}$, $L_{12}$ is proportional the streaming conductance $S_{\rm str}=\frac{\pi R^2}{\ell}L_{12}$, $L_{13}$ is proportional to the diffusio-osmotic mobility $D_{\rm DO}=k_{\rm B}TL_{13}$, $L_{22}$ is the electric conductivity of the channel and $L_{23}$ the diffusio-osmotic conductivity.

\begin{figure*}[!ht]
\centering
\includegraphics[width=0.95\textwidth]{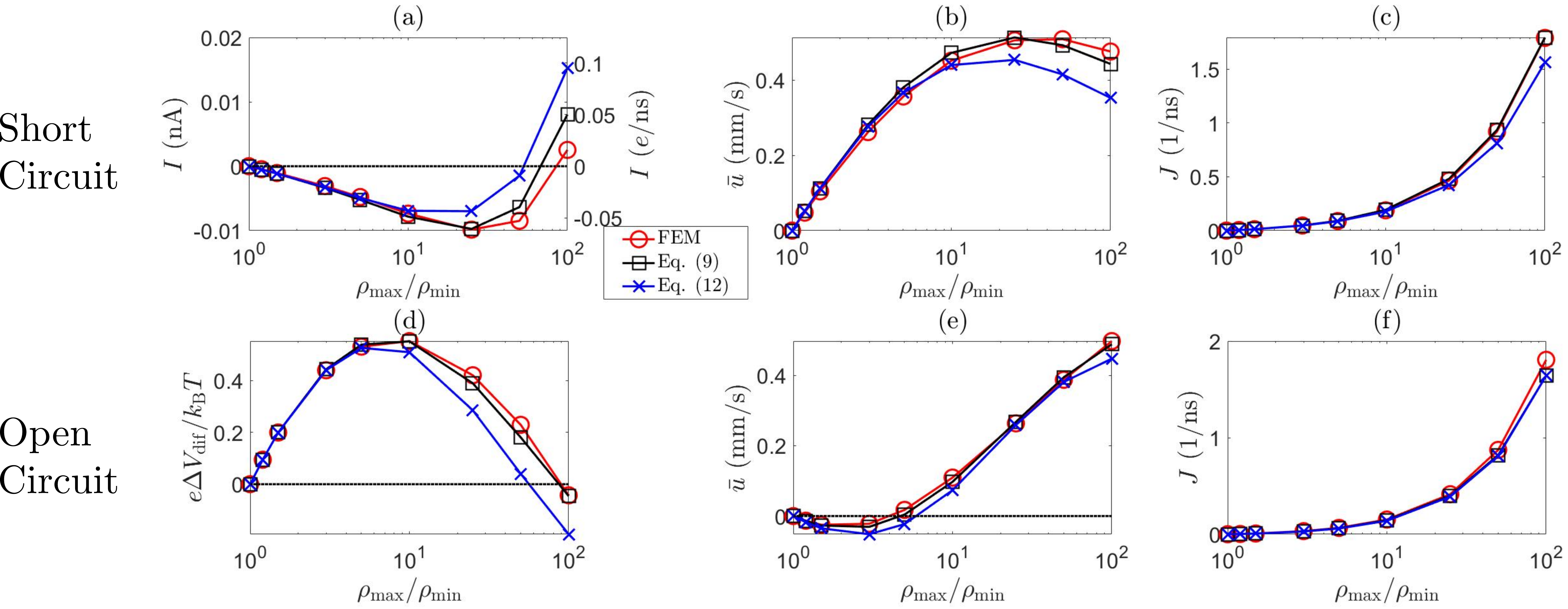}
\caption{The short-circuit electric current $I$, open-circuit potential $\Delta V_{\rm dif}$, average fluid velocity $\bar{u}=\frac{Q}{\pi R^2}$ and salt flux $J$ as a function of $\rho_{\rm max}/\rho_{\rm min}$. The red line represents the FEM results, the blue line the prediction of  \eqref{eq:Gana} and the black line \eqref{eq:cond} for $\sigma(\rho_s=1 {\rm mM})=-0.05$ $e$/nm$^2$ (\eqref{eq:CR}), $R=$ 60 nm, $b$=0, $\beta=-0.21$ (NaCl) and $\rho_{\rm min}=1$ mM for short-circuited (a)-(c) and open-circuit (d)-(f) system.}\label{fig:TestOns}
\end{figure*}

\section{Validation conductivity matrix}

Now that we have set up a formalism to extend the microscopic theory, represented by ${\bf L}$, to the global electrokinetic properties, represented by ${\bf G}$, we can compare the predictions of Eq. \eqref{eq:cond} and Eq. \eqref{eq:Gana} with the FEM solutions of the Nernst-Planck equations \eqref{eq:Stokesch3}-\eqref{eq:PNP} calculated using COMSOL Multiphysics, in order to validate the applicability of ${\bf G}$ via  Eq. \eqref{eq:cond} and Eq. \eqref{eq:Gana}. Here we will only focus on the diffusio-osmosis, as this inevitably includes significant lateral heterogeneities, for both a short-circuit and an open-circuit system as discussed above (Eq. \eqref{eq:IDO} and Eq. \eqref{eq:Vdif}).


Fig. \ref{fig:TestOns} shows the dependence of the average fluid velocity $\bar{u}=Q/(\pi R^2)$, electric current $I$ and salt flux $J$ on $\rho_{\rm max}/\rho_{\rm min}\in [1,100]$, with $\rho_{\rm min}=1$ mM, for NaCl from the FEM calculations compared to the predictions of  Eq. \eqref{eq:Gana} (blue) and  Eq. \eqref{eq:cond} (black), both for a short-circuit (a)-(c) and an open-circuit (d)-(f) system, for a charge regulation boundary condition with $\sigma(\rho_s=1 {\rm mM})=-0.05$ $e$/nm$^2$ (Eq. \eqref{eq:CR} with pH-pK=0.05), $b=0$ nm, $R=60$ nm, $\ell=1.5$ $\mu$m, $\rho_{\rm min}=1$ mM, $D_{\rm Na}=1.33\times 10^{-9}$ m$^2$/s and $D_{\rm Cl}=2.03\times 10^{-9}$ m$^2$/s ($\beta=-0.21$). Fig. \ref{fig:TestOns} shows that Eq. \eqref{eq:cond} is very accurate in reproducing the FEM results. In all cases, Eq. \eqref{eq:Gana} is less accurate than Eq. \eqref{eq:cond} but often surprisingly accurate given its simplifications, especially if $\rho_{\rm max}/\rho_{\rm min}\lesssim 10$ in both short-circuit and open-circuit conditions. The agreement in the open-circuit case thus shows that, even if there are multiple driving forces (i.e. both $\Delta V\neq0$ and $\Delta\mu\neq 0$) the formalism remains accurate. We have furthermore compared the predictions and the FEM calculations for a non-zero slip length ($b=10$ nm), smaller radius ($R=40$ nm), higher surface charge ($\sigma=-0.1$ e/nm$^2$), smaller channel length ($\ell=375$ nm) and higher minimum salinity ($\rho_{\rm min}=20$ mM) and found good agreement for all parameter variations (see Supplementary Information). In addition, Fig. \ref{fig:TestOns}(d) shows that in an open-circuit system $\Delta V_{\rm dif}$ changes sign for large $\Delta\mu$ since $I_{\rm DO}$ changes sign in the short-circuit case ($I_{\rm DO}$ changes sign due to the competition between $L_{23}^{\rm surf}$ and $L_{23}^{\rm vol}$). Moreover, Fig. \ref{fig:TestOns}(e) shows that the fluid flow first decreases, than increases and even changes sign with increasing $\rho_{\rm max}/\rho_{\rm min}$. This is the result of an intricate balance between diffusio-osmosis due to $\Delta\mu$ and electro-osmosis due to $\Delta V_{\rm dif}$. The balance between the diffusio-osmotic and electro-osmotic driving forces depends strongly on $\beta$, and is thus very different for KCl ($\beta=0$) than for NaCl ($\beta=-0.21$), and additionally depends on $z_{\sigma}$. Both of these behaviours are in agreement with experimental observations and interpretations \cite{Yang2014, He2016}.

\begin{figure}[!ht]
\centering
\includegraphics[width=0.48\textwidth]{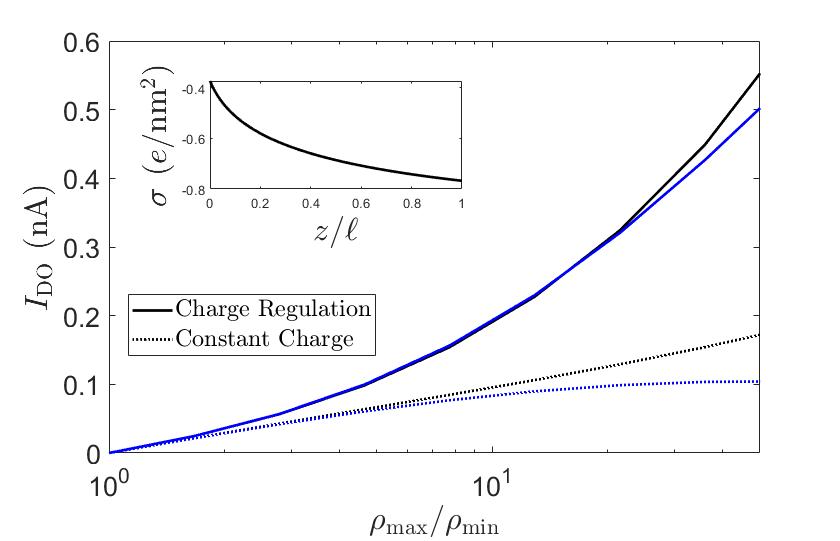}
\caption{The diffusio-osmotic current $I_{\rm DO}$ for KCl as a function of the salinity drop over the channel according to  Eq. \eqref{eq:Gana} (blue) and  Eq. \eqref{eq:cond} (black) for both a constant charge (CC, dashed) and charge regulation (CR, full) boundary condition. For both CC and CR, $\sigma(\rho_s=1 {\rm mM})=-0.25\, e$/nm$^2$, $\rho_{\rm min}=1$ mM, $R=40$ nm, $b=3$ nm and $\ell=1250$ nm. The inset shows the surface charge as a function of the lateral position $z$.}\label{fig:Ido}
\end{figure}

Recent experimental advances allow for direct comparison between theory and experiments for these kind of systems. For instance, measurements on osmotic power generation using a single boron nitride nanotubes (BNNT), carbon nanotubes (CNT) and MoS$_2$ nanopores, have been shown to surpass older RED technologies based on much thicker membranes \cite{Veerman2009}. With the theory presented here, we can directly compare with recent experiments. Fig. \ref{fig:Ido} shows the (short-circuit) diffusio-osmotic current $I_{\rm DO}$, for both a constant charge and a charge regulating boundary condition Eq. \eqref{eq:CR}, as a function of the salinity ratio $\rho_{\rm max}/\rho_{\rm min}$ for a nanochannel with $\rho_{\rm min}=1$ mM, $\sigma(\rho_s=1 {\rm mM})=-0.25$ e/nm$^2$, $R=40$ nm, $b=3$ nm and $\ell=1250$ nm, which can be directly compared to the diffusio-osmotic current measurements on BNNT by Siria et al. \cite{Siria2013}. Here, $\sigma(\rho_s=1 {\rm mM})$ was chosen such that similar $I_{\rm DO}$ values were obtained. First of all, it is evident from  Fig. \ref{fig:Ido} that, especially for large $\rho_{\rm max}/\rho_{\rm min}$, the charge regulation boundary condition has a significant effect on the predicted electric current. A charge regulation boundary condition (Eq. \eqref{eq:CR}) and the small slip length $b=3$ nm of BNNTs \cite{Tocci2014} are sufficient to obtain very similar values for $I_{\rm DO}$ (order 0.1 to 1 nA), but with a surface charge more than an order of magnitude smaller than estimated by Siria et al. \cite{Siria2013}. Note that the contribution from the slip length, which, for large $\sigma$, scales with $\sigma^2$ (see  Eq. \eqref{eq:Onscoef} and associated text), becomes increasingly dominant for increasing $\sigma$ (but was taken $b=0$ in by Siria et al. \cite{Siria2013}). Even a relatively small slip length of $b=3$ nm can therefore significantly affect the predicted fluxes. Note furthermore that $\sigma$ varies significantly as a function of the channel position $z$, see inset Fig. \ref{fig:Ido}, which explains why the charge regulation boundary condition gives a larger $I_{\rm DO}$ compared to the constant charge boundary condition, and furthermore emphasises the importance of even a small but finite $b$. 

The surface charge $\sigma(\rho_s=1$ mM$)=-0.25\,e/$nm$^2$ is much smaller that the value obtained from conductivity measurements on BNNT by Siria et al. \cite{Siria2013}. It has recently been shown, however, that the adsorbed OH$^-$ contributes significantly to the conductivity and other properties of the channel \cite{Grosjean2019}. Conduction via the Stern layer is not included in the current model, but an increased conduction will probably only lower the predicted surface charge even more. 
We have recently developed models for mobile surface charges \cite{Werkhoven2018,Werkhoven2019}, and incorporating these in the current theory is subject of future research.

\section{Reverse Electrodialysis}

Having established the accuracy of the theoretical framework of deriving ${\bf G}$ from ${\bf L}$, we can use the derived equations to analyse the wide variety of different electrokinetic systems (Table \ref{tab:systems}) without the need for full FEM calculations (or other extensive numerical analyses) for each system separately. All electrokinetic systems are essentially described by ${\bf G}$, the only difference being the boundary conditions. As an example, we will use the conductivity matrix ${\bf G}$ to analyse a single channel using Reverse Electrodialysis (RED), which are essentially intermediate between a short-circuit and open-circuit system. 

\begin{figure}[!ht]
\centering
\includegraphics[width=0.4\textwidth]{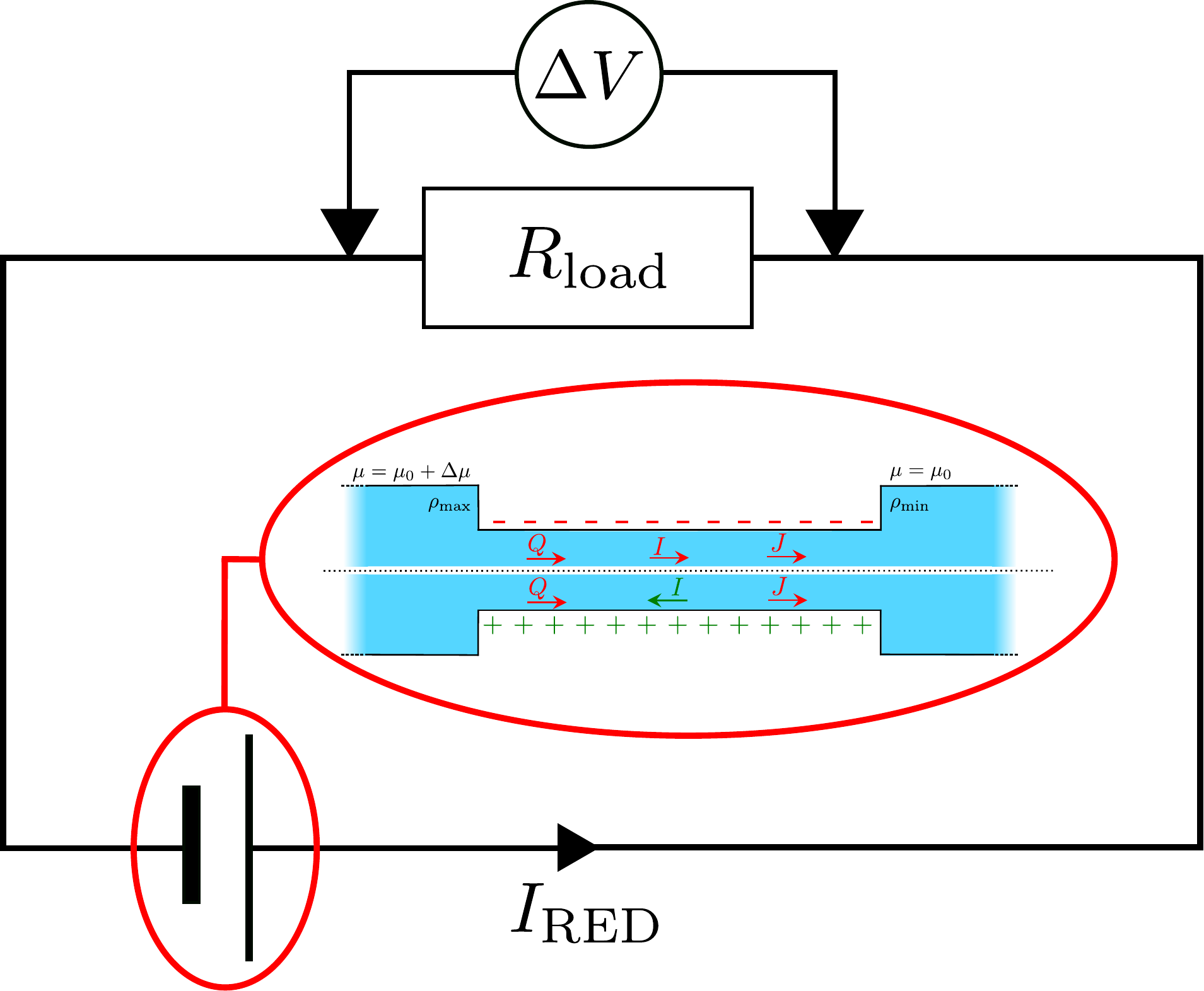}
\caption{Schematic representation of an RED circuit, where a diffusio-osmotic system is embedded in an electric circuit with a resistance $R_s$.}
\label{fig:RED}
\end{figure}

The electrokinetic RED system, schematically represented in Fig. \ref{fig:RED}, is embedded in an electric circuit and thus allows a non-zero current $I=I_{\rm RED}$ to flow through the system. However, the circuit also contains an (Ohmic) resistance $R_s$ that harvests the electric energy, which requires a potential drop $\Delta V$ in order for a non-zero current  to flow. Assuming that $R_s$ can be chosen freely, we will assume that $R_s$ is chosen such that the generated electric power is optimised (as opposed to the energy conversion efficiency). It is straightforward to show that the generated power is maximised when $R_s$ equals the resistance of the channel $R_{ch}=\frac{\ell}{\pi R^2 G_{22}}$ 
\cite{Kim2010,He2016}, which fixes the current to half the short-circuit current  Eq. \eqref{eq:IDO},
\begin{equation}\label{eq:circ}
I_{\rm RED}=\frac{1}{2}I_{\rm DO}=\frac{1}{2}\frac{\pi R^2}{\ell}G_{23}\Delta\mu_{\rm ch},
\end{equation}
with $I_{\rm DO}$ the short-circuit current, Eq. \eqref{eq:IDO}. Note that the resulting potential over the channel $\Delta V=IR_s$ is half the open-circuit (diffusion) potential $\Delta V_{\rm dif}$,  Eq. \eqref{eq:Vdif}, and that we must use $\Delta\mu_{\rm ch}$, the chemical potential drop over the channel, instead of $\Delta \mu$ to determine $I_{\rm DO}$. This allows us to write the maximum generated areal power density $P_{\rm RED}$ as
\begin{equation}\label{eq:power}
P_{\rm RED}=\dfrac{\mathcal{P}_{\rm RED}}{\pi R^2}=\dfrac{I_{\rm DO}^2R_{ch}}{4 \pi R^2}=\frac{1}{4}\frac{(\Delta\mu_{\rm ch})^2}{\ell}\frac{G_{23}^2}{G_{22}},
\end{equation}
where $\mathcal{P}_{\rm RED}$ is the generated electric power.  Eq. \eqref{eq:power} shows that the power density is inversely proportional to the length $\ell$, which (partially) explains the potential of nanopores 
\cite{Feng2016} compared to nanochannels, let alone microchannels. A smaller length decreases $R_{ch}$ (and $R_s$ is decreased accordingly) but increases the salinity gradient and thus $I_{\rm DO}$. The energy conversion efficiency can be found by dividing the generated electrical power by the osmotic free energy dissipated by the mixing of the two solutions 
\cite{Fair1971,Peters2016},
\begin{equation}\label{eq:eff}
\alpha_{\rm RED}=\dfrac{\mathcal{P}_{\rm RED}}{J_{\rm exc}\Delta\mu}=\dfrac{\mathcal{P}_{\rm RED}}{J\Delta\mu-2\Delta\rho Q},
\end{equation}
see Supplementary Information for a derivation why $\alpha_{\rm RED}$ is defined with $J_{\rm exc}$ and $\Delta\mu$ instead of $J$ and/or $\Delta\mu_{\rm ch}$. Whether it is "better" to maximise the power or the efficiency depends on the goal and the available resources. In the case of diffusio-osmosis both fresh and salt water are available in abundance where rivers flow into the sea, so it makes sense to optimise for the generated power. A similar analysis can be performed for mechanical energy conversion, where a pressure drop $\Delta p$ is used to generate an electric current (via $G_{12}$), but osmotic energy converters have been shown to be able to produce more energy at a higher conversion efficiency
\cite{vanderHeyden2006,Lu2006}. 

On the basis of Eq. \eqref{eq:power} and Eq. \eqref{eq:eff}, we are in the position to use the conductivity matrix ${\bf G}$ to investigate the effect of system parameters on the RED performance without the need to run intensive FEM or other numerical calculations for each parameter set. As mentioned, two materials have shown great potential for osmotic energy conversion: CNTs and BNNTs. The reason for the success of the former is believed to be related to the small friction of water with the surface, i.e. a large slip length $b$, 
\cite{Majumder2005, Pang2011, Tocci2014, Marbach2016}, while for the latter the large surface charge is believed to be main cause 
\cite{Siria2013}, in addition to the large conductivities shown by both 
\cite{Pang2011, Siria2013,Secchi2016,Grosjean2019}. For both materials, we assume a charge regulating boundary condition as in  Eq. \eqref{eq:CR}. There are many parameters to investigate, but here we will focus on 4 main aspects: the surface charge density $\sigma$, the slip length $b$, the minimum salt concentration $\rho_{\rm min}$ and the mobility mismatch $\beta$. 

As an example we will investigate a nanochannel with $R=40$ nm, although it should be kept in mind that for RED a smaller $R$ generally results in a higher $P_{\rm RED}$ and $\alpha_{\rm RED}$. However, the slip length of CNTs is known to vary with $R$ \cite{Marbach2016}, so a constant $R$ allows us to assume a constant $b$ for this analysis. We will use $b_{\rm BNNT}= 3$ nm as the slip length for BNNTs \cite{Tocci2014} and $b_{\rm CNT}= 30$ nm for the slip length of CNTs \cite{Tocci2014, Marbach2016}.


\begin{figure}[!ht]
\centering
\hspace{-0.5cm}
\includegraphics[width=0.5\textwidth]{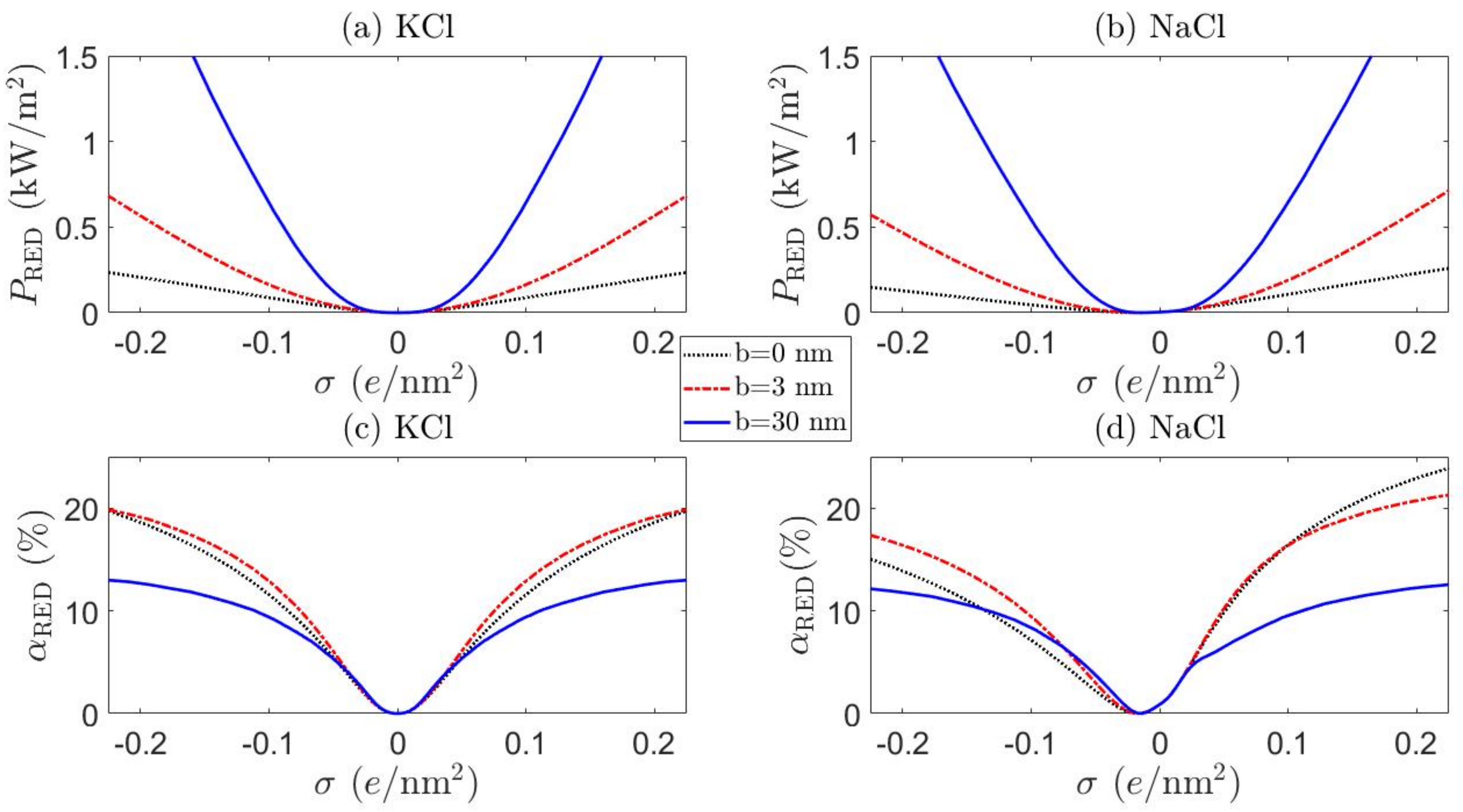}
\caption{The RED generated power $P_{\rm RED}$ (a) \& (b) and efficiency $\alpha$ (c) \& (d) for KCl (a) \& (c) ($\beta=0$) and NaCl (b) \& (d) ($\beta=-0.21$) as a function of the surface charge at $\rho_s=1$ mM, for channel lengths $\ell=1.5 \,\mu$m, $\rho_{\rm max}=25$ mM, $\rho_{\rm min}=1$ mM (so $\Delta\mu=k_{\rm B}T\log 25$) and radius $R=40$ nm ($R\approx 4\lambda_{D, {\rm min}}$). The dotted black line represent $b=0$, the red dashed line represents $b= 3$ nm (BNNT) and the full blue line represents $b=30$ nm (CNT).}\label{fig:REDlowsal}
\end{figure}

\begin{figure}[!ht]
\centering
\hspace{-0.5cm}
\includegraphics[width=0.5\textwidth]{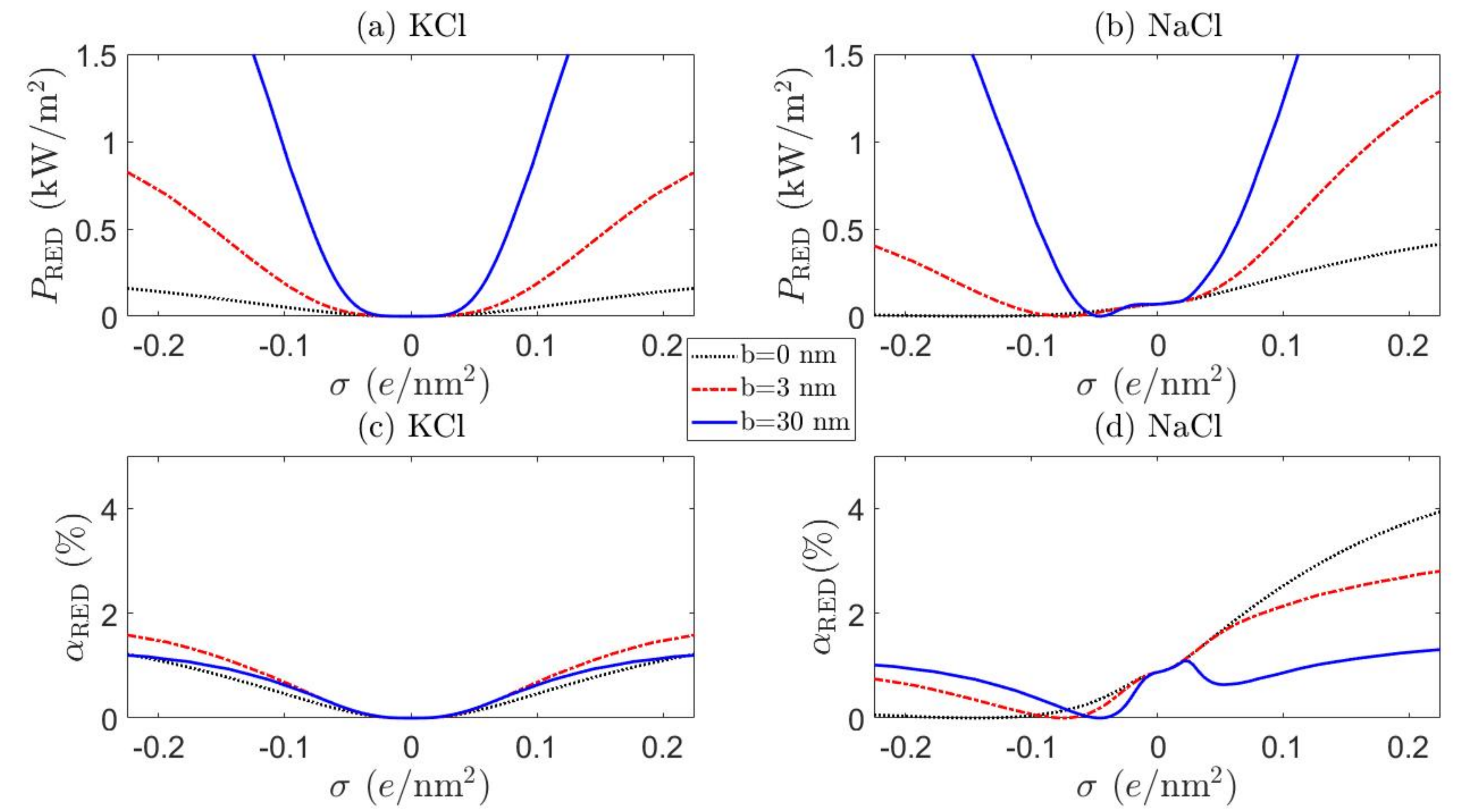}
\caption{As in caption Fig. \ref{fig:REDlowsal}, but with $\rho_{\rm max}=500$ mM and $\rho_{\rm min}=20$ mM.}\label{fig:REDhighsal}
\end{figure}

Fig. \ref{fig:REDlowsal} and Fig. \ref{fig:REDhighsal} show the RED power and efficiency for $\rho_{\rm min}=1$ mM and $\rho_{\rm min}=20$ mM, respectively, for both KCl (a) \& (c) ($\beta=0$) and NaCl (b) \& (d) ($\beta=-0.21$), for $b=0$ (black dotted), $b=3$ nm (red dot-dashed) and $b=30$ nm (blue full) as a function of  $\sigma$. The horizontal axis represents the surface charge $\sigma$ at $\rho_s=1$ mM. The surface charge of both BNNT and CNT surfaces originate from an OH$^-$ adsorption reaction \cite{Siria2013, Secchi2016, Grosjean2019} and strictly only takes negative values. Positive values are included (H$^+$ adsorption), however, for a more complete analysis. There are a few observation we can make from these figures.

First of all, a comparison of the black ($b=0$), red (BNNT, $b=3$ nm) and blue (CNT, $b=30$ nm) shows that not only a large but also a moderate slip length $b$ has a significant effect on the electrokinetic properties of the system, as was also noted for mechanical energy conversion 
\cite{Ren2008}. This confirms that the large slip length of CNTs makes these nanochannels so promising. In addition, Fig. \ref{fig:REDlowsal} confirms the point emphasised above, that even a small $b$ can have significant effects on the current through the channel, especially for large $\sigma$.

Secondly, we see that the predicted power can significantly differ between KCl and NaCl, especially for large $\rho_{\rm min}$, shown in Fig. \ref{fig:REDhighsal}.  Many experiments are performed with KCl, but it is not a priori clear whether these results can be extrapolated to NaCl (the main species of salt for large-scale applications of RED). The difference between these two salts originates from the mobility mismatch, $\beta_{\rm KCl}\approx 0$ and $\beta_{\rm NaCl}\approx -0.21$, which not only affects the resulting fluxes but also breaks the charge inversion symmetry (see Eq. \eqref{eq:Onscoef}). 
If NaCl is the main constituent of the electrolyte, a positively charged surface is more effective than a negatively charged surface: a negatively charged surface will attract the cations to the surface, but Na$^+$ has a lower mobility than Cl$^-$. The EDL thus has a lower overall mobility if $\sigma<0$ than if $\sigma>0$. This provides a general rule that RED systems generate more power at a higher efficiency if $z_{\sigma}\beta<0$, because then the ion with the highest mobility is the most abundant in the EDL. 

A comparison of Fig. \ref{fig:REDlowsal} and Fig. \ref{fig:REDhighsal} furthermore emphasises the point that the generated power and efficiency do not purely depend on the concentration ratio (i.e. $\Delta\mu$), but are both a function of the the separate salinities $\rho_{\rm min}$ and $\rho_{\rm max}$ 
\cite{Kim2010}. This is especially true for NaCl, for which the broken inversion symmetry is significantly more apparent for $\rho_{\rm min}=20$ mM (Fig. \ref{fig:REDhighsal}) than for $\rho_{\rm min}=1$ mM (Fig. \ref{fig:REDlowsal}). Especially if $b=0$, the difference between the two cases is very pronounced (compare black dotted line Fig. \ref{fig:REDlowsal}(b) \& (d) and Fig. \ref{fig:REDhighsal}(b) \& (d)). The dependence on $\rho_{\rm min}$ can be understood by the fact that $L^{\rm vol}_{23}$, and consequently $G_{23}$ and $I_{\rm DO}$, increase with $\beta\rho_s$ (see Eq. \eqref{eq:Onscoef}). All slip-length contributions are, however, independent of $\rho_s$, and all scale as $b\sigma^2$ for large $\sigma$ (see  Eq. \eqref{eq:Onscoef}). 

We also find that the generated power for $\rho_{\rm min}=20$ mM and $\rho_{\rm max}=500$ mM is higher than for $\rho_{\rm min}=1$ mM and $\rho_{\rm max}=25$ mM if $b=0$, especially for NaCl with $\sigma>0$. However, the efficiency $\alpha_{\rm RED}$ is nearly an order of magnitude higher for $\rho_{\rm min}=1$ mM than for $\rho_{\rm min}=20$ mM, even though the chemical potential drop $\Delta\mu$ is the same in both cases. Both can be understood by the increased role played by the volume contributions ${\bf L}^{\rm vol}$ of Eq. \eqref{eq:Onscoef}. These contributions scale with $\rho_s$, so an increased $\rho_{\rm min}$ naturally leads to a larger $I_{\rm DO}$ (if $\beta\neq 0$ via $L_{23}^{\rm vol}$) and thus a larger $P_{\rm RED}$. Similarly, the total salt flux $J$ increases with $\rho_{\rm min}$ ($L^{\rm vol}_{33}\propto \rho_s$) which, in turn, decreases $\alpha_{\rm RED}$ (see  Eq. \eqref{eq:eff}). 

Finally, note that $P_{\rm RED}$ and $\alpha_{\rm RED}$ develop a minimum, with a minimum value of zero, for NaCl with a small negative surface charge. This minimum shifts to larger values of $\sigma$ if $\rho_{\rm min}$ increases, since this minimum is given by the value of $\sigma$ for which the volume and surface contributions to $I$ cancel. If we take the surface charge of CNTs at $\rho_s=1$ mM to be $\sigma=-(0.03-0.1)\, e$/nm$^2$ \cite{Pang2011,Marbach2016}, we even find that CNT are typically not far removed from the minimum observed in Fig. \ref{fig:REDhighsal}. We should note, however, that the location of this minimum depends on systems parameters such as $R$ and $b$, so this does not mean that CNTs should not be used for RED. It does, on the other hand, stress the important point that $\beta$, $\sigma$ (including its sign) and $\rho_{\rm min}$ are important parameters to keep in mind when optimising a given channel.


Note that our values for $P_{\rm RED}$ are of the same order of magnitude as measurements on BNNTs 
\cite{Siria2013}. These values are also consistent with measurements on nanopores 
\cite{Feng2016}, where they found $P_{\rm RED}$ three orders of magnitude higher than for micron-thick membranes, with $\ell$ three orders of magnitude lower. The predictions do certainly depend on the radius $R$, as RED typically generates more power per unit area and is more efficient for smaller $R$ 
\cite{Kim2010}. The present analysis, however, emphasis the point that different systems with differing $R$, $\rho_{\rm min}$, $\sigma$ (including its sign), $b$ and $\beta$, are optimised differently. There is of course an immense variety when it comes to nanochannels, but the framework presented in this article provides an accessible method with which these channels can be analysed. Moreover, the framework can be further improved, for example for smaller $R$, because the most restricting assumption of the Onsager matrix presented in this article, Eq. \eqref{eq:Onscoef}, is the assumption of non/weakly-overlapping EDLs, meaning that  Eq. \eqref{eq:Onscoef} is viable for $R\gtrsim 12$ nm for $\rho_{\rm min}>10$ mM. There is no general analytic theory for the matrix elements of ${\bf L}$ for arbitrary $\lambda_D/R$, but it is possible to take the thin-pore limit ($\lambda_D\ll R$) of the Poisson-Boltzmann formalism to obtain analytical solutions \cite{Bocquet2010}. In addition, the Poisson-Boltzmann formalism typically breaks down for $\rho_s>100$ mM, but there are theories to improve on Poisson-Boltzmann 
\cite{Borukhov1997,Bazant2019}. Lastly, as already stated, it has been shown that surface conduction plays an important role for BNNTs and CNTs \cite{Grosjean2019}, which can further affect the (quantitative) predictions of the theory. This will be the subject of future research.

\section{Summary \& conclusion}

In conclusion, we have presented a method to fully analyse the transport properties of electrokinetic channels driven by a pressure gradient, an electric field or a salinity gradient. We have calculated the full 3$\times$3 Onsager matrix ${\bf L}$ which gives the volumetric flow rate, electric current and salt flux for a given (set of) driving force(s), which to be best of our knowledge was absent in the current literature. This includes an important contribution to the diffusio-osmotic electric current that has so-far been overlooked. We then presented two methods to extend the local linear-response Onsager matrix ${\bf L}$ to a global linear-response conductivity matrix ${\bf G}$, which can incorporate lateral heterogeneities. This furthermore allowed us to include more complex boundary conditions such as charge regulation boundary condition. We compared the predictions of the theory with numerically exact (Finite Element Method) solutions of the Poisson-Nernst-Planck-Stokes equations, which showed the remarkable accuracy of the theory under varying parameters and boundary conditions. Charge regulation was shown to have a significant effect on the predicted fluxes, and thus on the interpretation of recent experiment on nanochannels. 

Having established the accuracy of the conductivity matrix ${\bf G}$, we used it to analyse Reverse Electrodialysis without the need to use extensive numerical calculations such as FEM. We compared typical values for Carbon Nanotubes and Boron Nitride Nanotubes, and showed, for example, that such systems behave differently when KCl is used compared to NaCl. Most notably, in the case of NaCl we showed that negatively charged surfaces such CNTs and BNNTs are significantly less effective than positively charged surfaces, especially if salinities like those of fresh and sea water are used. We furthermore emphasised that the produced power does not solely depend on the chemical potential drop across the channel, but on the reservoir salinities separately. We thus found that systems with different surfaces charge, different type of salt and salinities are optimised differently. Electrokinetic systems present a very large parameter space, too large to fully explore here, but for this reason electrokinetic systems represent a great variability and applicability. The framework presented in this article provides an insightful and convenient method to analyse them.


\section*{Acknowledgements}
This work is part of the D-ITP consortium, a program of the Netherlands Organisation for Scientific Research (NWO) that is funded by the Dutch Ministry of Education, Culture and Science (OCW). This work is a part of an Industrial Partnership Program of the Netherlands Organization for Scientific Research (NWO) through FOM Concept agreement FOM-15-0521. Financial support was provided through the Exploratory Research (ExploRe) programme of BP plc.



\bibliographystyle{apsrev4-1}

\clearpage

\appendix
\begin{center}
\textbf{\Large{Supplementary Information}}
\end{center}

\section{Dissipation \& symmetry Onsager Matrix}

In general, we can define an Onsager matrix between a set of fluxes and associated driving forces. However, as mentioned in the main text, the Onsager matrix is only symmetric if the driving force and associated flux are congruent, i.e. that the dissipation rate is given by the product of the flux and the driving force \cite{Onsager1,Onsager2}. We can write the dissipation rate $T\dot{S}$ as \cite{Fair1971}
\begin{equation}
T\dot{S}=-\sum\limits_{i=0}^2j_i\Delta\nu_i,
\end{equation}
where $\Delta\nu_i$ is the total electrochemical potential difference of the $i$th species between the two reservoirs and $i=0$ for the solvent, $i=1$ for the cation and $i=2$ for the anion. We can write down the electrochemical potential of the ions as
\begin{equation}
\Delta\nu_i=v_i\Delta p+\Delta\mu_i+z_i e\Delta V,
\end{equation}
with $v_i$, $\rho_i$ and $z_i$ the volume of a particle, the density and the valency of species $i$, $\Delta p$ the pressure drop, $\Delta V$ the voltage drop and $\Delta\mu_i=k_{\rm B}T\Delta(\log\rho_i)$ the chemical potential drop across the channel. Note that $\Delta\mu_1=\Delta\mu_2$ due to the charge neutrality of the reservoirs. We assume the solvent to be incompressible, and therefore we can write the partial solvent pressure $\Delta p_0$ as
\begin{equation}
\Delta\nu_0=v_0\Delta p_0.
\end{equation}
Now we can use van 't Hoffs law to write the total pressure $p$ as
\begin{equation}
p=p_0+\Pi=p_0+2\rho_sk_{\rm B}T,
\end{equation}
with $\Pi$ the partial solute pressure. Note that in equilibrium, $p$ is constant even if $\Pi$ is not. Now we can write the dissipation rate as
\begin{gather}
\begin{aligned}
T\dot{S}&=-j_0v_0\Delta p_0-(j_1v_1+j_2v_2)\Delta p-J\Delta\mu-I\Delta V,\\
&=-Q\Delta p+j_0v_0\Delta\pi-J\Delta\mu-I\Delta V,
\end{aligned}
\end{gather}
where we have defined the volume flux $Q=j_0v_0+j_1v_1+j_2v_2$, solute or salt flux $J=j_1+j_2$, charge flux $I=e(j_1-j_2)$ and chemical potential drop $\Delta\mu=k_{\rm B}T\Delta(\log\rho_s)$ (equal for both ions due to charge neutrality in the bulk). For dilute solutions we have that $Q\approx j_0v_0$, and we can rewrite
\begin{equation}
\begin{aligned}
T\dot{S}&=-Q\Delta p+2k_{\rm B}TQ\Delta\rho_s-J\Delta\mu-I\Delta V\\
&=-Q\Delta p-\left(J-2k_{\rm B}TQ\dfrac{\Delta\rho_s}{\Delta\mu}\right)\Delta\mu-I\Delta V,\\
&=-Q\Delta p-J_{\rm exc}\Delta\mu-I\Delta V,
\end{aligned}
\end{equation}
where we have identified the excess salt flux $J_{\rm exc}$
\begin{equation}\label{eq:Jexc}
J_{\rm exc}=J-2k_{\rm B}TQ\dfrac{\Delta\rho_s}{\Delta\mu}.
\end{equation}
In order for ${\bf L}$ to be symmetric, $J_{\rm exc}$ is congruent to $\Delta\mu$. Additionally, \eqref{eq:Jexc} shows how to obtain the total salt flux $J$ from the excess salt flux $J_{\rm exc}$ even if $\Delta \mu\neq 0$. Note that
\begin{equation}
\lim\limits_{\Delta\rho_s\rightarrow 0} k_{\rm B}T\dfrac{\Delta \rho_s}{\Delta\mu}=\lim\limits_{\Delta\rho_s\rightarrow 0} k_{\rm B}T\dfrac{\Delta\rho_s}{\log\left(1+\frac{\Delta\rho_s}{\rho_1}\right)}=\rho_s,
\end{equation}
with $\Delta\rho_s=\rho_2-\rho_1$ the salinity drop over the channel. \eqref{eq:Jexc} is therefore also valid if $\Delta\mu=0$. 


\section{Derivation ${\bf G}^{\rm vol}$}

The volume contribtions of the Onsager matrix are given by
\begin{equation}
{\bf L}^{\rm vol}=\begin{pmatrix} L_{11} & 0 & 0 \\ 0 & m e^2\rho_s & m e\beta\rho_s \\ 2\rho_sL_{11} & 2( m e \beta+ L_{12}) \rho_s & 2(m +L_{13})\rho_s \end{pmatrix},
\end{equation}
with $m=\frac{D_++D_-}{2k_{\rm B}T}$ the salt mobility, with the inverse
\begin{equation}
\left({\bf L}^{\rm vol}\right)^{-1}=\begin{pmatrix} \dfrac{1}{L_{11}} & 0 & 0 \\ -\dfrac{\beta}{\Delta} & -\dfrac{m+L_{13}}{2m e \Delta}\dfrac{1}{\rho(z)} & \dfrac{\beta}{2 \Delta} \dfrac{1}{\rho(z)}\\ \dfrac{e}{\Delta} & \dfrac{\beta(me+L_{12})}{2m e \Delta}\dfrac{1}{\rho(z)} & -\dfrac{e}{2 \Delta}\dfrac{1}{\rho(z)} \end{pmatrix},
\end{equation}
with $\Delta=\beta^2(m e +L_{12})-e(m+L_{13})$ a constant. Given a linear $\rho_s(z)=\rho_1+\frac{z}{\ell}\Delta\rho$, with $\Delta \rho=\rho_2-\rho_1$, we have that
\begin{equation}
\int\limits_0^{\ell}{\rm d}z\dfrac{1}{\rho_s(z)}=\ell \dfrac{\log\frac{\rho_2}{\rho_1}}{\Delta\rho},
\end{equation}
and we find, with $\Delta\mu=k_{\rm B}T\log\frac{\rho_2}{\rho_1}$, that
\begin{gather}\label{eq:Gvol}
\begin{aligned}
{\bf G}^{\rm vol}&=\dfrac{1}{\ell}\left(\int\limits_0^{\ell}{\rm d}z{\bf L}^{-1}(\sigma=0)\right)^{-1}\\
&=\begin{pmatrix} L_{11} & 0 & 0 \\ 0 & 2De^2\frac{\Delta\rho}{\Delta\mu} & 2De\frac{\Delta\rho}{\Delta\mu}\beta \\ \frac{k_{\rm B}T\Delta\rho}{\Delta\mu}L_{11} & 2(D e+k_{\rm B}TL_{12})\frac{\Delta\rho}{\Delta\mu}\beta & 2(D+k_{\rm B}TL_{13})\frac{\Delta\rho}{\Delta\mu} \end{pmatrix}
\end{aligned}
\raisetag{2.0cm}
\end{gather}

\section{Entrance effects}

As mentioned in the main text, the salinity at both ends of the channel are not equal to the salinities imposed on the bulk, $\rho_{\rm max}$ and $\rho_{\rm min}$. The effect is not necessarily strong, but a small change in especially the low salinity can have a significant effect on the (local) conductivity. It is therefore important to take these entrance effects into account, and the predictions are indeed much more accurate if we do. We cannot solve for the concentration profile exactly (due to the complicated fluid flow en electrostatic potential profile), but we can get a good estimate by assuming that the concentration profile outside the channel drops off over a typical distance $R$. Since the diffusion equation has no intrinsic length scale, the geometric length $R$ should characterise the concentration gradients outside the channel. Therefore we approximate
\begin{equation}\label{eq:entr}
\begin{aligned}
\rho_{\rm out}&\approx\rho_{\rm min}+R\partial_z\rho_s,\\
\rho_{\rm in}&\approx\rho_{\rm max}-R\partial_z\rho_s,\\
\end{aligned}
\end{equation}
where $\rho_{\rm out}$ is the salinity at the outlet (low salinity side) and $\rho_{\rm in}$ the salinity at the entrance (high salinity side). Note that the salinity gradient must be expressed in terms of $\rho_{\rm out}$ and $\rho_{\rm in}$, $\partial_z\rho_s=\frac{\rho_{\rm out}-\rho_{\rm in}}{\ell}$ which we can plug into \eqref{eq:entr} and solve for $\rho_{\rm out}$ and $\rho_{\rm in}$ to find
\begin{equation}
\begin{aligned}
\rho_{\rm out}&\approx\rho_{\rm min}+\frac{R}{\ell+2R}\Delta \rho,\\
\rho_{\rm in}&\approx\rho_{\rm max}-\frac{R}{\ell+2R}\Delta \rho,
\end{aligned}
\end{equation}
where $\Delta\rho=\rho_{\rm max}-\rho_{\rm min}$ is the imposed salinity drop across the channel. 
\begin{figure}[!ht]
\centering
\includegraphics[width=0.45\textwidth]{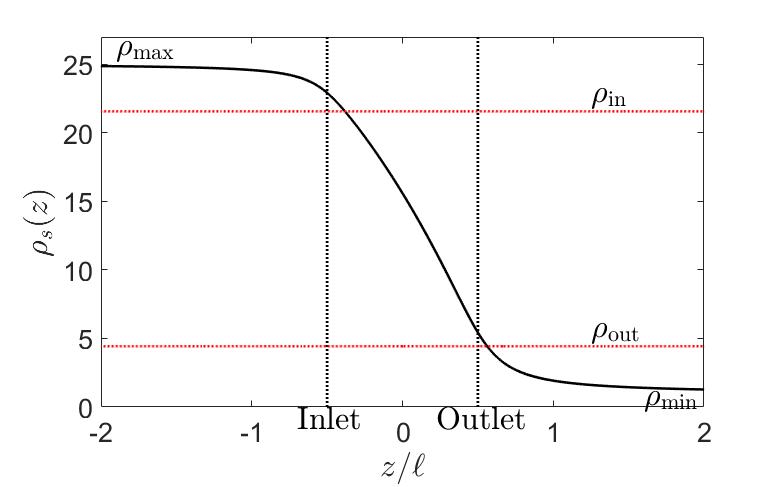}
\caption{Density profile at the axis of the channel calculated with FEM (black full line) for KCl, $D_{\rm K^+}\approx D_{\rm Cl^-}=2\times 10^{-9}$ m$^2$/s, $R=60$ nm and $\ell=300$ nm. The dashed red lines indicate the inlet and outlet salinity $\rho_{\rm in}\approx 22$ mM and $\rho_{\rm out}\approx 4$ mM, and the black dashed lines indicate the location of the inlet ($z=-\frac{1}{2}\ell$) and outlet ($z=-\frac{1}{2}\ell$).}\label{fig:sal2}
\end{figure}
As has been shown in the main text, the entrance effect are relevant even for needle-shaped channels. As the aspect ratio increases, however, the entrance effects become even stronger. For example, Fig. \ref{fig:sal2} shows the entrance effects for a channel with $R=60$ nm and $\ell=300$ nm. Here we see that $\rho_{\rm in}$ is almost a factor 5 larger than $\rho_{\rm min}$, significantly affecting the total conductivity.

\section{Validation theory: parameter variation}

Below we will show the validation of the presented theories under several parameter variations, for the diffusio-osmotic current $I_{\rm DO}$, average fluid velocity $\bar{u}=\frac{Q_{\rm DO}}{\pi R^2}$ and salt flux $J_{\rm DO}$. The red line represents the FEM results, the blue line the analytic approach (Eq. \ref{eq:Gana}), the black line exact approach (Eq. \ref{eq:cond}). The numerical uncertainty of $I_{\rm DO}$ increases with $\rho_{\rm max}/\rho_{\rm min}$, and is typically of the order of a few pico Amp\`eres for $\rho_{\rm max}/\rho_{\rm min}=25$, i.e. typically much smaller than the size of the symbols. The figures below use the parameter set $\sigma=-0.05$ $e$/nm$^2$, $R=$ 60 nm, $\ell=1.5\, \mu$m, $b$=0 nm, and $\rho_{\rm min}=1$ mM with a constant charge boundary condition, but with every figure one exception stated in the caption.
\begin{figure}[!ht]
\centering
\includegraphics[width=0.47\textwidth]{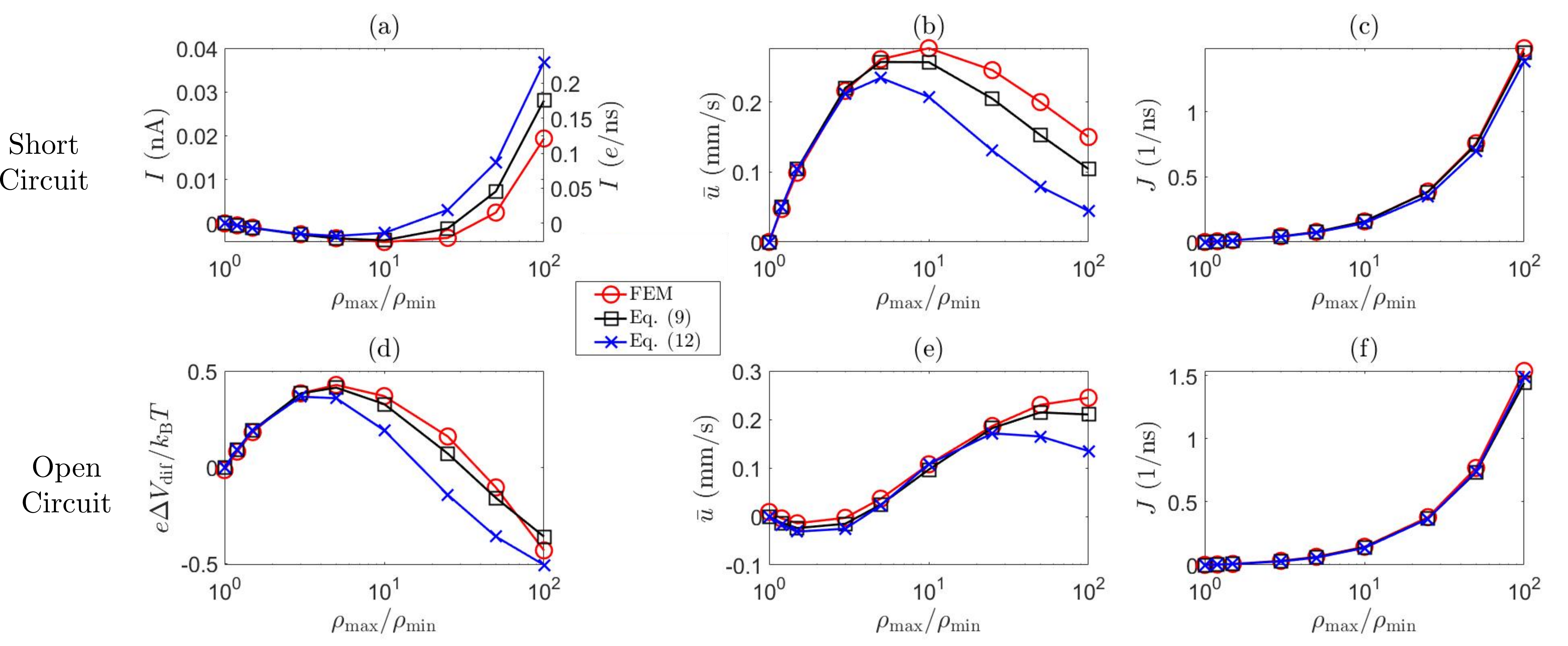}
\caption{NaCl (a)-(c) KCl (d)-(f), with a charge regulation boundary condition \eqref{eq:CR} and the parameter set stated in the text.}
\end{figure}
\begin{figure}[!ht]
\centering
\includegraphics[width=0.47\textwidth]{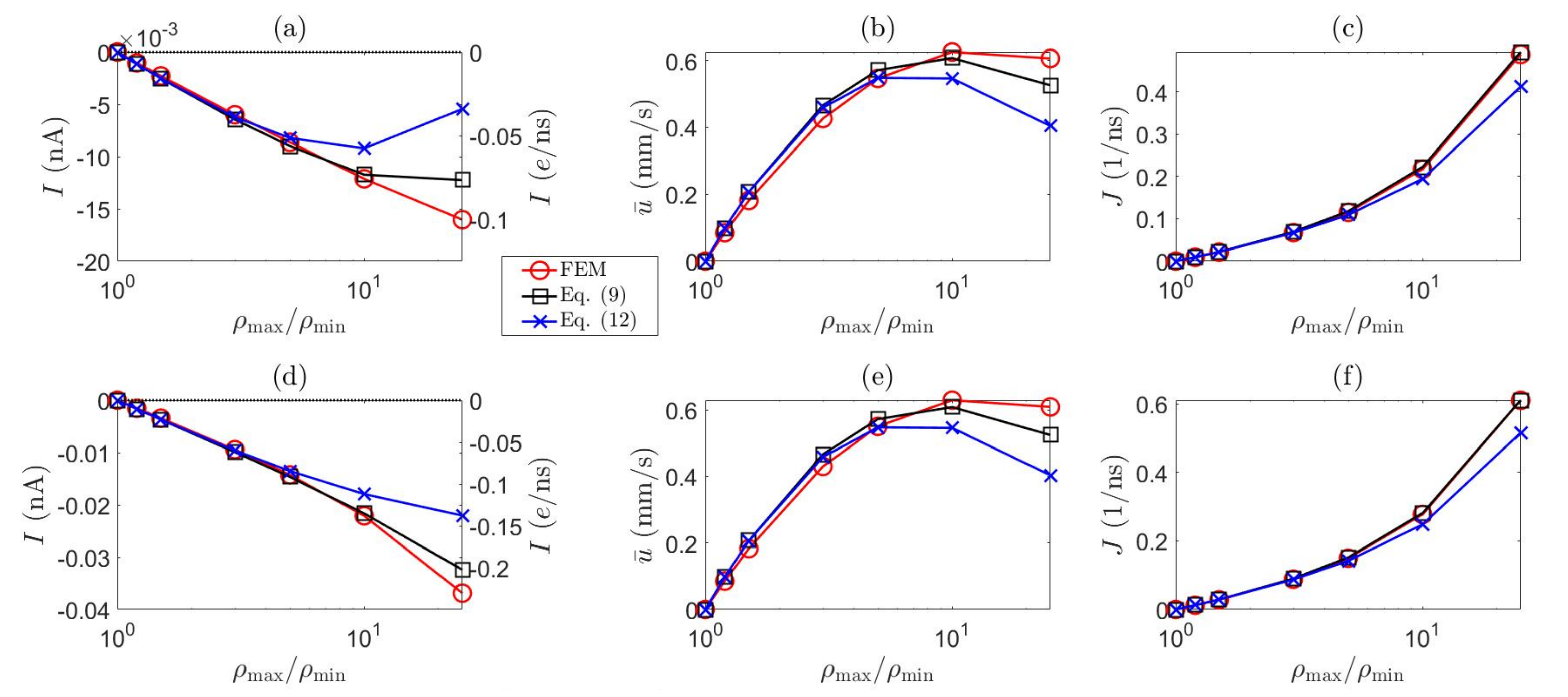}
\caption{NaCl (a)-(c) KCl (d)-(f), with $\sigma=-0.1$ $e$/nm$^2$ and the parameter set stated in the text.}
\end{figure}
\begin{figure}[!ht]
\centering
\includegraphics[width=0.47\textwidth]{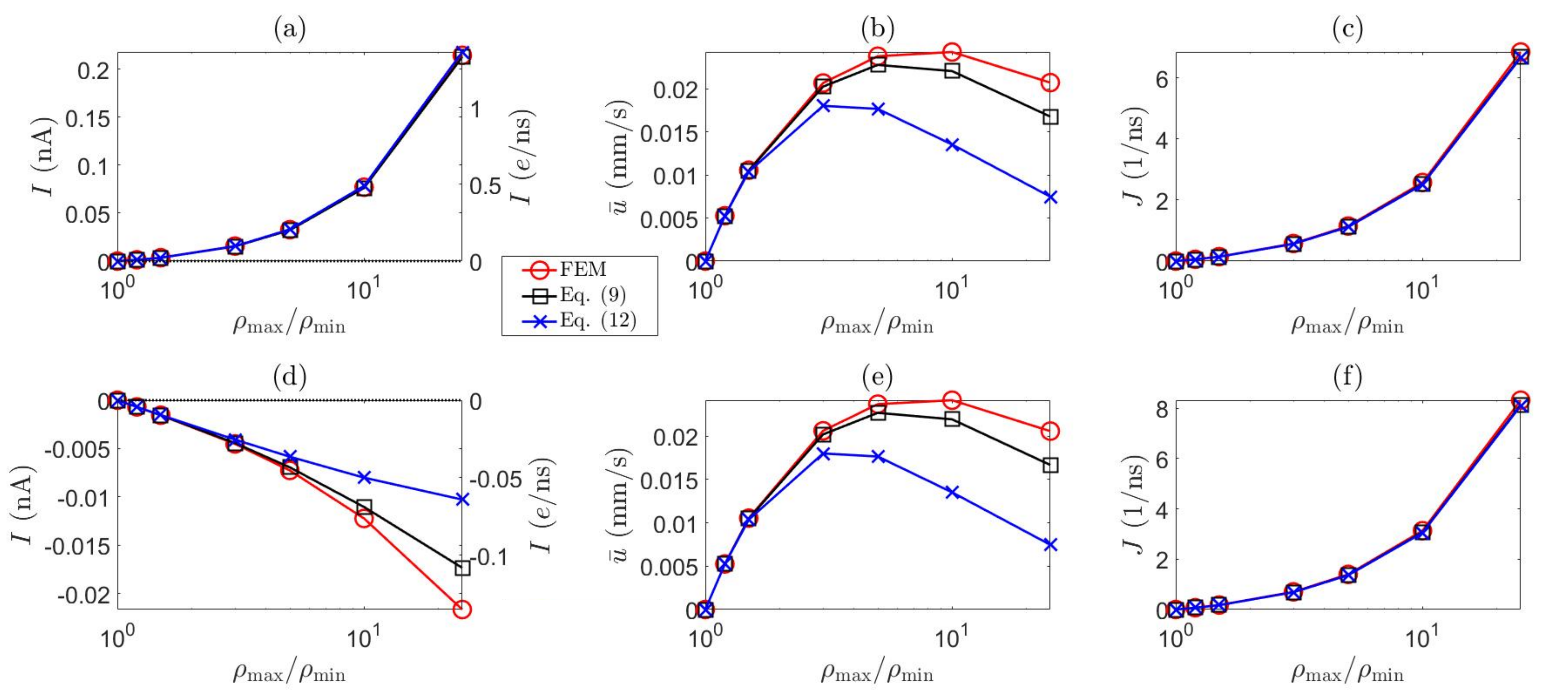}
\caption{NaCl (a)-(c) KCl (d)-(f), with $\rho_{\rm min}=20$ mM and the parameter set stated in the text.}
\end{figure}
\begin{figure}[!ht]
\centering
\includegraphics[width=0.47\textwidth]{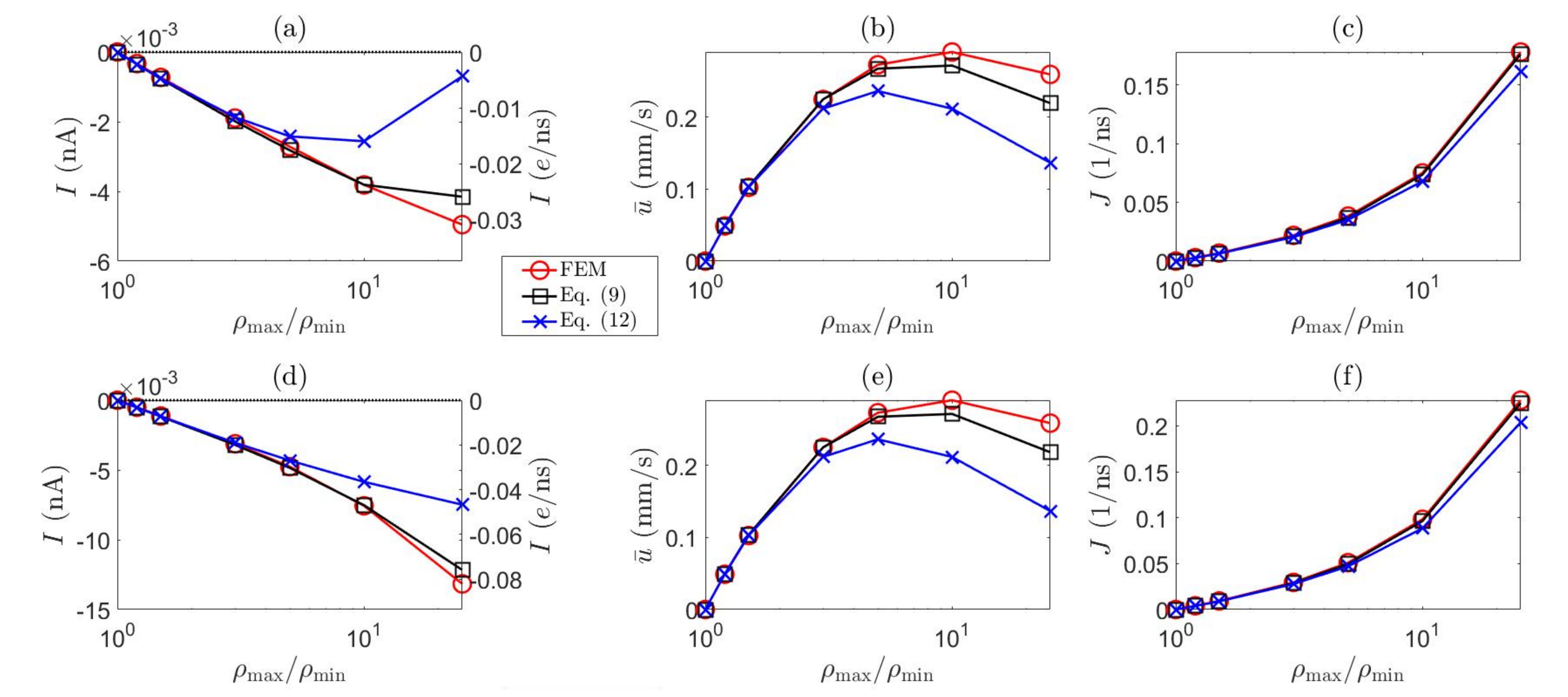}
\caption{NaCl (a)-(c) KCl (d)-(f), with $R=$ 40 nm and the parameter set stated in the text.}
\end{figure}
\begin{figure}[!ht]
\centering
\includegraphics[width=0.47\textwidth]{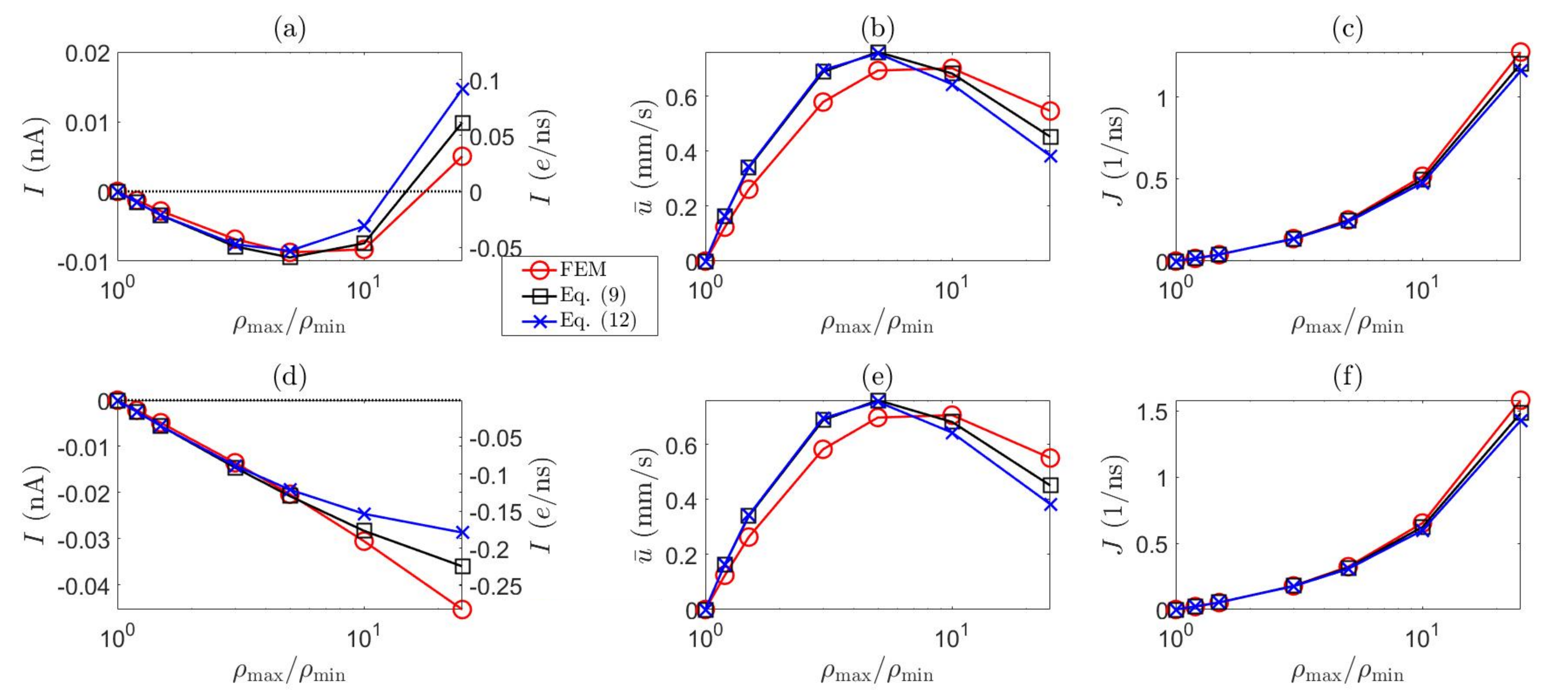}
\caption{NaCl (a)-(c) KCl (d)-(f), with $\ell=0.375\, \mu$m and the parameter set stated in the text.}
\end{figure}
\begin{figure}[!ht]
\centering
\includegraphics[width=0.47\textwidth]{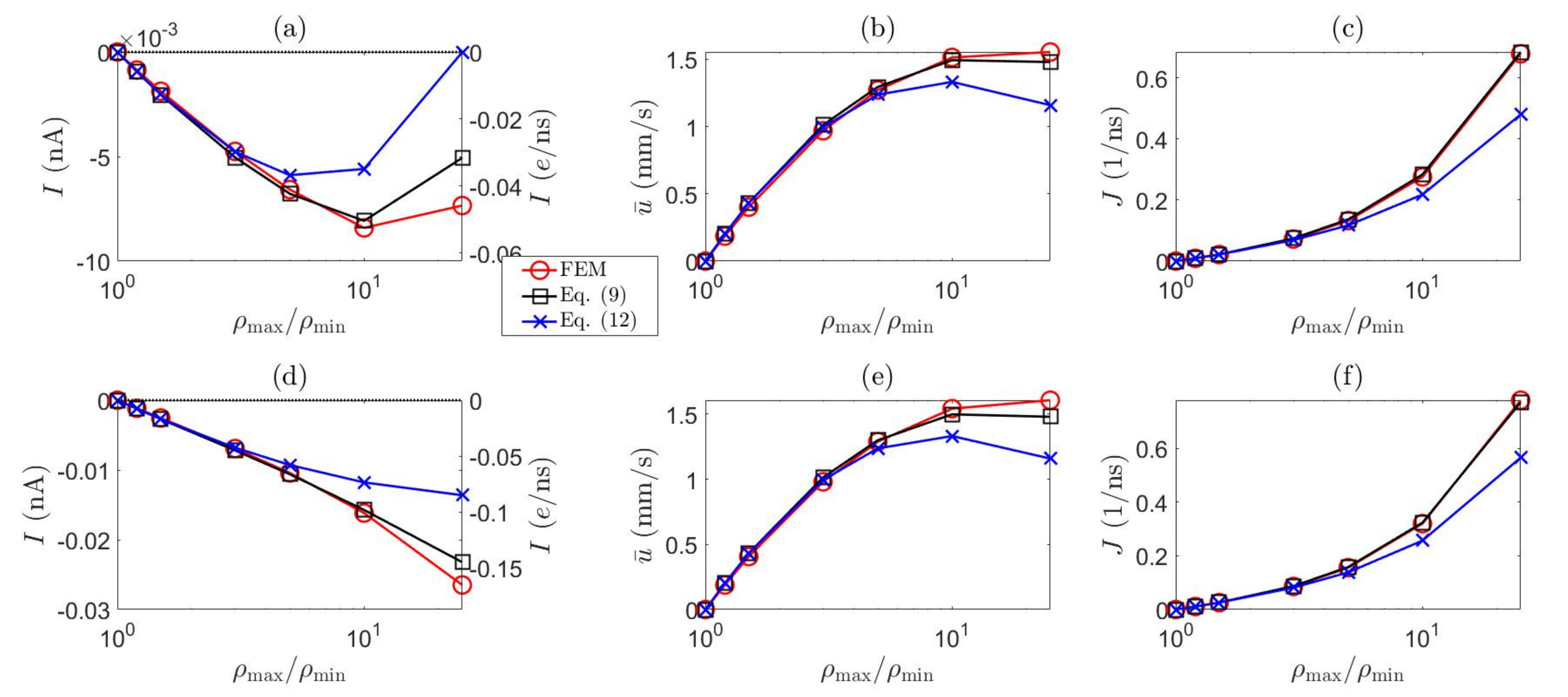}
\caption{NaCl (a)-(c) KCl (d)-(f), with $b$=10 nm, and $\rho_{\rm min}=1$ mM and the parameter set stated in the text.}
\end{figure}

\newpage

\section{COMSOL simulation}

\begin{figure}[!ht]
\centering
\includegraphics[width=0.49\textwidth]{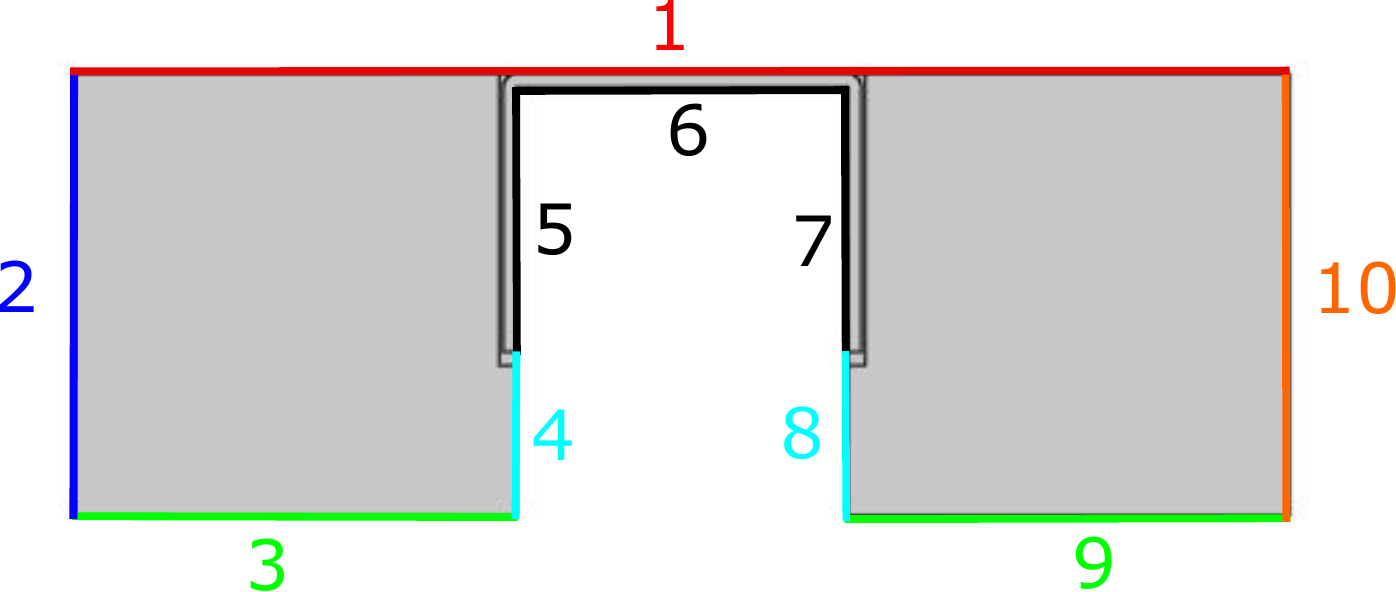}
\caption{Domain on which the governing equations are solved numerically with the boundaries marked (boundary conditions explained in the text.}\label{fig:domain}
\end{figure}

For each simulation domain use the boundary conditions.
\begin{enumerate}
\item[1] ({\color{red} Red}) Axis of rotational symmetry. All normal derivatives and velocities are zero, ${\bf n}\cdot \nabla \rho_i=0={\bf n}\cdot \nabla \psi={\bf n}\cdot {\bf u}$
\item[2] ({\color{blue} Dark blue}) Inlet reservoir, where we fix the pressure $p=\Delta p$, salinities $\rho_{\pm}=\rho_{\rm max}$ and potential $\psi=\Delta V$
\item[3,9] ({\color{green} Green}) To simulate an infinite bulk, we impose no-slip boundary conditions $u_z=0$ on the side of the bulk if $\Delta p\neq 0$, and otherwise an open boundary (force free boundary), and impose a fixed salinity ($\rho_{\pm}=\rho_{\max}$ for 3 and $\rho_{\pm}=\rho_{\rm min}$ for 9).
\item[4,8] ({\color{cyan} Cyan}) Hard walls with slip boundary condition ${\bf n_s \cdot \nabla} {\bf u}_t=b{\bf u}_t$ (with ${\bf u}_t$ the tangential component of the velocity), zero-charge $\mathbf{n}\cdot \nabla\psi=0$ and no-flux boundary conditions, $\mathbf{n}\cdot \mathbf{J}_i=0$
\item[5,6,7] (\textbf{Black}) The charged wall. The same boundary conditions as 4/8 except for a charged boundary condition $\mathbf{n}\cdot \nabla\psi=-\frac{\sigma}{\epsilon}$, with $\epsilon$ the permittivity and the surface charge $\sigma$ determined by the wished boundary condition (constant charge, charge regulation etc.).
\item[10] ({\color{orange} Orange}) Outlet reservoir, $\rho_{\pm}=\rho_{\rm min}$, $\psi=0$ and an open-boundary condition for the fluid.
\end{enumerate}

\section{The Onsager matrix} 

\subsection{Poisson-Boltzmann identities}

For the calculation of ${\bf L}$ we assume that channel radius $R$, is significantly larger than the Debye length. This allows us to significantly simplify the equations in cylindrical coordinates for quantities evaluated close to the surface. In this case, we make a coordinate transformation $s=R-r$ such that
\begin{equation}
\begin{aligned}
&\nabla^2 f=\frac{1}{r}\frac{\partial}{\partial r}\left(r\frac{\partial f}{\partial r}\right)\approx \frac{\partial^2 f}{\partial s^2}, \\ 
&\int\limits^R_0 \dif r 2 \pi r f(r)\approx 2 \pi R\int\limits_0^R \dif s f(s),
\end{aligned}
\end{equation}
for any function $f(r)$ that only takes non-zero values inside the EDL. Therefore, if we are only considering quantities inside the EDL all calculations are basically the same if we consider a cylinder or parallel plate geometry, except for a prefactor. The parallel plate expressions can be found by simply substituting $\pi R\rightarrow H$, with $H$ the plate separation. The error for the cylindrical geometry is of the order of $\lambda_D/R$, but for the parallel plate geometry the only error occurs as soon as the EDL significantly overlap. The expressions for ${\bf L}$ presented below assume non-overlapping EDLs. However, the theory remains accurate even for weakly-overlapping EDLs, since in that case the density profiles and electrostatic potential are very well approximated by the sum of the individual EDLs. 

First we give the equilibrium Gouy-Chapmann expressions for a 1:1 salt \cite{Schoh2008,overbeek} that we assume to hold for the electric double layer, where $\kappa=\lambda_D^{-1}$ is the inverse Debye length, $\sigma$ the density of surface charges, $z_{\sigma}$ the sign of the surface charge, $\phi_0$ the dimensionless surface potential and $\sigma^{\ast}=(2\pi\lambda_B\lambda_D)^{-1}$.
\begin{equation}\label{eq:PB}
\begin{aligned}
\phi(s)&=4\,{\rm arctanh}(\gamma e^{\kappa s})=2\log\frac{1+\gamma e^{-\kappa s}}{1-\gamma e^{-\kappa s}};\\
\gamma&= \tanh\frac{1}{4}\phi_0=\frac{\sigma^{\ast}}{\sigma}\left(\sqrt{1+\left(\frac{\sigma}{\sigma^{\ast}}\right)^2}-1\right),\\
\rho_{\pm}&=\rho_s e^{\mp\phi};\qquad \frac{\sigma}{\sigma^{\ast}}=\frac{2\gamma}{1-\gamma^2}; \qquad 4\rho_s\lambda_D=\sigma^{\ast};\\
\sigma&=\sigma^{\ast}\sinh\frac{1}{2}\phi_0;\\
\sigma\gamma &= \sqrt{\sigma^{\ast 2}+\sigma^2}-\sigma^{\ast}=4\rho_s\lambda_D(\cosh\frac{1}{2}\phi_0-1).
\end{aligned}
\end{equation}
Next, we define a set of integrals as a function of the EDL potential which we encounter in the calculation of the Onsager coefficients. Each of these integrals are defined such they are positive, and each of these can be calculated analytically using the 1:1 Poisson-Boltzmann expressions, \eqref{eq:PB}.
\begin{gather} \label{eq:integrals}
\begin{aligned}
P_1= &\frac{z_{\sigma}}{\lambda_D}\int\limits_0^{\infty}\dif s \phi(s)=2\left({\rm Li}_2(|\gamma|)-{\rm Li}_2(-|\gamma|)\right),\\
P_2= &\frac{1}{\lambda_D}\int\limits_0^{\infty} \dif s (\cosh\phi(s)-1)=2(\cosh\frac{1}{2}\phi_0-1)\\
&=\frac{\sqrt{\sigma^{\ast 2}+\sigma^2}-\sigma^{\ast}}{2\rho_s\lambda_D},\\
P_3= &\frac{1}{\lambda_D^2}\int\limits_0^{\infty}\dif s s (\cosh\phi(s)-1)=4\log\cosh\frac{1}{4}\phi_0,\\
P_4= &\frac{1}{\lambda_D^3}\int\limits_0^{\infty}\dif s s^2 (\cosh\phi(s)-1)=2{\rm Li}_2(\gamma^2),\\
P_5=&-\frac{z_{\sigma}}{\lambda_D}\int\limits_0^{\infty}\dif s(\phi-\phi_0)(\cosh\phi-1)=\\
=&2z_{\sigma}\left(2\sinh\frac{1}{2}\phi_0-\phi_0\right)=4\frac{|\sigma|}{\sigma^{\ast}}-2|\phi_0|,\\
P_6=&-\frac{1}{\lambda_D}\int\limits_0^R\dif s (\cosh\phi-1)\log\left(1-\gamma^2 e^{-\kappa s}\right)\\
&=2P_3\cosh^2\frac{1}{4}\phi_0-P_2,\\
P_3P_2&-2P_6=2P_2-4P_3.\\
\end{aligned}
\raisetag{5cm}
\end{gather}
The first integral $P_1$ can be solved by rewriting $\phi(s)$ in terms of the polylogarithmic function Li$_1$, and for the integrals $P_2$-$P_6$ we can use the Poisson-Boltzmann identities
\begin{equation}\label{eq:trick}
\begin{aligned}
&\phi(s)=2\log\frac{1+\gamma e^{-\kappa s}}{1-\gamma e^{-\kappa s}}=-{\rm Li}_1\left(\gamma e^{-\kappa s}\right)+{\rm Li}_1\left(-\gamma e^{-\kappa s}\right),\\
&2(\cosh\phi-1)=\dfrac{4\gamma e^{-\kappa s}}{\left(1-\gamma e^{-\kappa s}\right)^2}-\dfrac{4\gamma e^{-\kappa s}}{\left(1+\gamma e^{-\kappa s}\right)^2}.
\end{aligned}
\end{equation}

\subsection{Calculation $L_{11}$}

Poiseuille flow through a cylindrical channel is given by
\begin{equation}\label{eq:Stokessol}
u_z(r)=-\frac{\partial_z p}{4\eta}\left(R^2-r^2+2Rb\right)
\end{equation}
This allows us to find the volumetric flow rate and thus the first Onsager coefficient
\begin{equation}
L_{11}=\frac{1}{\pi R^2}\frac{Q_{\rm S}}{-\partial_z p}=\frac{\ell}{\pi R^2 \Delta p}\int\limits_0^R\dif r 2\pi r u(r)=-\frac{R^2}{8\eta}\left(1+\frac{4b}{R}\right),
\end{equation}
where $\partial_zp=-\Delta p/\ell$.

\subsection{Calculation $L_{12}$}

The generated charge current due to fluid flow is given by
\begin{equation}
I_S=2\pi e\int\limits_0^r\dif rr \rho_e(r)u(r).
\end{equation}
We use the Poisson equation to eliminate $\rho_e$, and then use partial integration twice to obtain
\begin{equation}
\begin{aligned}
I_S=&-2\pi R\epsilon\left(-\partial_s\psi(0) u(0) +\psi(0) \partial_s u(0)+\int\limits_0^{R}\dif s \psi \partial_s^2 u \right),\\
&=-2\pi R\epsilon\frac{\partial_z p}{4\eta}\left(-\frac{e\sigma}{\epsilon}2bR-2R\psi_0 +z_{\sigma}\frac{k_{\rm B}T\lambda_D}{e}P_1\right).
\end{aligned}
\end{equation}
Here we used Gauss' law $\epsilon\partial_s\psi(0)=-\sigma$, with $\sigma$ the areal density of surface charges, and that $\partial_s^2u=\partial_zp/(4\eta)$ from \eqref{eq:Stokessol}. Additionally, we defined $\psi_0=\psi(0)$ as the surface potential and $P_1$ is one of the Poisson-Boltzmann integrals defined above (Eq. \ref{eq:integrals}). Note that $P_1$ is a positive, dimensionless number which is still a function of the surface charge. We can now write the next Onsager coefficient as
\begin{equation}
L_{12}=\frac{1}{\pi R^2}\frac{I_{S}}{-\partial_z p}=-\frac{\epsilon\psi_0+be\sigma}{\eta}+z_{\sigma}\frac{e\lambda_D}{2\pi\lambda_B \eta R}P_1,
\end{equation}

\subsection{Calculation $L_{31}$}

Lastly, we must calculate the ion flux $J_{\rm S, exc}=J_s-2\rho_sQ_S$, given by
\begin{equation}
\begin{aligned}
J_{\rm exc, S}&=J_{\rm S}-2\rho_s Q_{\rm S}=2\pi\int\limits_0^R\dif r r(\rho_++\rho_-2\rho_s)u\\
&=2\pi\int\limits_0^R\dif r r(\rho_++\rho_--2\rho_s)u.
\end{aligned}
\end{equation}
This integral can now straightforwardly be rewritten as
\begin{equation}
J_{\rm exc, S}=\dfrac{\pi\rho_s\partial_z p}{\eta}R\int\limits_0^R\dif s (\cosh\phi(s)-1)\left(2R(s+b)-s^2\right).
\end{equation}
In order to calculate $J_{\rm S, exc}$, we need three Poisson-Boltzmann integrals $P_2$, $P_3$ and $P_4$ defined above (\eqref{eq:integrals}), such that we can write the next Onsager coefficient can thus be expressed as 
\begin{equation}\label{eq:L13}
L_{13}=\frac{1}{\pi R^2}\frac{J_{\rm exc, S}}{-\partial_z p}=\frac{1}{4\pi\lambda_B\eta}\left(\frac{b}{\lambda_D}P_2+P_3-\frac{\lambda_D}{2R}P_4\right).
\end{equation}

\subsection{Calculation $L_{12}$}

In the electrically driven case we have no externally applied pressure gradient and Stokes' equation reduces to
\begin{equation}
\eta\nabla^2u+\rho_e E=0,
\end{equation}
where $E=\Delta V/\ell$ is the applied electric field, $\Delta V$ the applied potential drop over the channel and $\ell$ the length of the channel. Substituting Poisson's equation we find
\begin{equation}
\partial^2_z u=\frac{\epsilon E}{\eta}\partial^2_z\psi.
\end{equation}
This equation can be integrated twice to give
\begin{equation}\label{eq:EOFlow}
u(s)=u_0+\frac{\epsilon E}{\eta}(\psi(s)-\psi_0)=\frac{E}{\eta}\left(\epsilon(\psi(s)-\psi_0)-be\sigma\right).
\end{equation}
Now we can calculate electro-osmotic volumetric flow rate $Q_{\rm EO}$,
\begin{equation}
Q_{\rm EO}=2\pi\int\limits^R_0\dif r r u(r)=\pi R^2 u_{\rm EO}+2\pi R\int\limits^R_0\dif s (u(r)-u_{\rm EO}),
\end{equation}
here, $u_{\rm EO}$ is the electro-osmotic fluid flow, the (constant) fluid velocity outside of the EDL, and we have used that $u-u_{\rm EO}$ is only non-zero in the EDL. Now we find
\begin{equation}
\begin{aligned}
Q_{\rm EO}&=2\pi\int\limits^R_0\dif r r u(r)=-\pi R^2 E \frac{\epsilon\psi_0+be\sigma}{\eta}\\
&+2\pi R\frac{\epsilon E}{\eta}\frac{z_{\sigma}k_{\rm B}T}{e}\lambda_D P_1,
\end{aligned}
\end{equation}
and subsequently the next Onsager coefficient $L_{21}$,
\begin{equation}
L_{21}=\frac{1}{\pi R^2}\frac{Q_{\rm EO}}{E}=-\frac{be\sigma+\epsilon\psi_0}{\eta}+\frac{z_{\sigma}e\lambda_D}{2\pi\lambda_B  \eta R}P_1.
\end{equation}
Here we see that indeed $L_{12}=L_{21}$ as it should.

\subsection{Calculation $L_{22}$}

In the electrically driven case, we have both electric field and a fluid flow, so the current is composed of an advective ($I_{\rm EO, adv}$) and a conductive current ($I_{\rm EO, con}$). The conductive contribution to the current can expressed as
\begin{equation}
\begin{aligned}
I_{\rm EO, con}&=2\pi \frac{e^2}{k_{\rm B}T} E \int\limits_0^R \dif r r(D_+\rho_+(r)+D_-\rho_-(r)),\\
&=\pi \frac{e^2}{k_{\rm B}T}D\rho_s E\int\limits_0^R \dif r r(\cosh\phi-\beta\sinh\phi),
\end{aligned}
\end{equation}
where $D=\frac{1}{2}(D_++D_-)$, with $D_{\pm}$ the diffusion constant of the cation/anion and $\beta=\frac{D_+-D_-}{D_++D_-}$. We must be careful here, since the integrand is not only non-zero inside the EDL, so we should not simply change coordinates to $s$. Therefore we split the integral in a bulk and a surface contribution,
\begin{equation}\label{eq:IEOcond}
I_{\rm EO, con}=4\pi \frac{e^2}{k_{\rm B}T}D\rho_s E\left(\int\limits_0^R \dif r r+R\int\limits_0^R \dif s (\cosh\phi-1-\beta\sinh\phi)\right),
\end{equation}
where we have changed the coordinates from $r$ to $s$ in the second integral since the integrand is only non-zero inside the EDL. We can recognise $P_2$ in the second term on the right hand side, and the last term is easily determined using charge conservation condition
\begin{equation}\label{eq:neutr}
\int\limits_0^R\dif s(\rho_+-\rho_-)=-2\rho_s\int\limits_0^R\dif s\sinh\phi=-\sigma,
\end{equation}
This allows us to write down the conductive contribution to the current,
\begin{equation}
I_{\rm EO, cond}=4\pi R \frac{e^2}{k_{\rm B}T}D\rho_sE\left(\frac{1}{2}R+\lambda_D P_2-\frac{1}{2}R\beta \frac{\sigma}{\rho_s}\right).
\end{equation}
This leaves us to determine the advective contribution to the current $I_{\rm EO, adv}$ using \eqref{eq:EOFlow}
\begin{equation}
I_{\rm EO, adv}= 2\pi Re\frac{E}{\eta}\int\limits_0^R\dif s (\rho_+-\rho_-)(\epsilon(\psi(s)-\psi_0)-be\sigma).
\end{equation}
Interestingly, we find in $I_{\rm EO, adv}$ the self energy of the EDL, which can be expressed as
\begin{equation}
\frac{1}{2}e\int\limits_0^R\dif s(\rho_+-\rho_-)\psi=-\frac{1}{2}e\sigma\psi_0-\frac{k_{\rm B}T}{4\pi\lambda_B\lambda_D}P_2.
\end{equation}
Combining this with the charge neutrality condition used above, Eq. (\ref{eq:neutr}), we find
\begin{equation}
I_{\rm EO, adv}=2\pi Re^2\frac{E}{\eta}\left(b\sigma^2+\frac{2}{(4\pi\lambda_B)^2\lambda_D}P_2\right).
\end{equation}
Collecting all terms we find for the total electro-osmotically driven electric current and thus $L_{22}$
\begin{equation}\label{eq:IEO}
\begin{aligned}
L_{22}=&\frac{1}{\pi R^2}\frac{I_{\rm EO}}{E}\\
=&\frac{2De^2}{k_{\rm B}T} \left(\rho_s+\frac{2\rho_s\lambda_D}{R}P_2\left(1+\frac{k_{\rm B}T}{2\pi\lambda_B\eta D}\right)-\beta\frac{\sigma}{R} \right)\\
&+2\frac{b}{R}\frac{e^2\sigma^2}{\eta}
\end{aligned}
\end{equation}

\subsection{Calculation $L_{32}$}

Just like $I_{\rm EO}$, $J_{\rm EO}$ contains contributions from both conduction and advection. The conduction contribution can be calculated similar to $I_{\rm EO,cond}$,
\begin{gather}\label{eq:NEOcond}
\begin{aligned}
J_{\rm EO,cond}&=2\pi \frac{e}{k_{\rm B}T} \int\limits_0^R\dif r r (D_+\rho_+-D_-\rho_-)E\\
&=4\pi\beta D\rho_s eE\int\limits_0^R\dif r r (-\sinh\phi+\beta(\cosh\phi-1)+\beta)
\end{aligned}
\raisetag{1.35cm}
\end{gather}
We have already solved these equation above, \eqref{eq:IEOcond}, so here it suffices to state the result
\begin{equation}
J_{\rm EO,cond}=2\pi R\frac{e}{k_{\rm B}T}D E\left(-\sigma+\beta(\rho_sR+2\rho_s\lambda_DP_2)\right).
\end{equation}
We can find the advective contribution $J_{\rm EO, adv}$ as
\begin{equation}
\begin{aligned}
J_{\rm exc, EO, adv}&=2\pi R\int\limits_0^R\dif s (\rho_++\rho_--2\rho_s) u\\
&=4\pi R\frac{E}{\eta}\int\limits_0^R\dif s (\cosh\phi-1)(\epsilon(\psi-\psi_0)-be\sigma).
\end{aligned}
\end{equation}
This integral introduces yet another Poisson-Boltzmann identity $P_5$, see \eqref{eq:integrals}, and we find
\begin{equation}
J_{\rm exc, EO,adv}=-2\rho_s \pi R^2\frac{eE\lambda_D}{\eta R}\left(\frac{z_{\sigma}}{2\pi\lambda_B}P_5+2 b\sigma P_2\right).
\end{equation}
This gives the next Onsager coefficient
\begin{equation}
\begin{aligned}
L_{23}=&\frac{1}{\pi R^2}\frac{J_{\rm EO}}{E}=-\frac{2De}{k_{\rm B}T}\left(\frac{\sigma}{R}-\beta\rho_s\left(1+\frac{2\lambda_D}{R}P_2\right)\right)\\
&-\frac{e}{2\pi\lambda_B\lambda_D\eta R}\left(\frac{z_{\sigma}}{4\pi\lambda_B}P_5+b\sigma P_2\right)
\end{aligned}
\end{equation}

\subsection{Calculation $L_{13}$}

Contrary to an applied pressure or voltage difference, a concentration gradient does not directly induce a fluid flow because there is no body force directly related to the concentration gradient. In order for a concentration gradient to induce a fluid flow, an external potential is required that works in a direction perpendicular to the concentration gradient. In the case of a concentration gradient along a charged surface, this external potential is the electrostatic potential of the EDL. We will again assume that the EDL is in (local) equilibrium at every point along the surface. Since the salinity $\rho_s$ is a function of $z$, $\psi$ is a function of both $z$ as well as $r$. Interestingly, as we will see, the lateral electric field originating from $\psi(r,z)$ will not affect the resulting fluid flow profile. First, we write the ion densities as
\begin{equation}
\rho_{\pm}(r,z)=\rho_s(z)e^{\mp \phi(r,z)},
\end{equation}
where $\phi= \frac{e}{k_{\rm B}T} \psi$ is the dimensionless EDL potential. Assuming that the $r$ component of the fluid velocity vanishes, so we can write down the $r$ component of Stoke's equation
\begin{equation}
\partial_r p=-k_{\rm B}T\rho_0(z)\left(e^{-\phi}-e^{\phi}\right)\partial_r\phi=2k_{\rm B}T\rho_s(z)\partial_r\left(\cosh\phi\right).
\end{equation}
Now we can easily solve for pressure, and since the pressure must be constant ($p_0$) outside of the EDL (a concentration gradient cannot induce a fluid flow without the external potential) we find
\begin{equation}\label{eq:pDO}
p(r,z)=p_0+2k_{\rm B}T\rho_s(z)\left(\cosh\phi(r,z)-1\right).
\end{equation}
It is this pressure, which results from a concentration gradient through the EDL, which induces the fluid flow. Plugging \eqref{eq:pDO} in Stokes equation we find
\begin{equation}\label{eq:DifOsmStokes}
\begin{aligned}
\eta\partial^2_r u&=2k_{\rm B}T\partial_z\left(\rho_s(z)\left(\cosh\phi(r,z)-1\right)\right)-e\rho_e E_z\\
&=2k_{\rm B}T\partial_z\rho_s\left(\cosh\phi-1\right)+2k_{\rm B}T\rho_s\sinh\phi\partial_z\phi-e\rho_e(r) E_z\\
&=2k_{\rm B}T\partial_z\rho_s\left(\cosh\phi-1\right)-e\rho_e\partial_z\psi-e\rho_e E_z,\\
&=2k_{\rm B}T\partial_z\rho_s\left(\cosh\phi-1\right),
\end{aligned}
\end{equation}
where we defined $E_z=-\partial_z\psi$. Interestingly, this is the same result as the result we woudl obtain if we neglected the $z$ dependence of the EDL potential $\phi(r,z)=\phi(r)$, although we should keep in mind that $u$ is now a function of $z$ even in linear response theory. 

It is possible to find an exact solution to this equation with the Poisson-Boltzmann formalism. To solve for the diffusio-osmotic flow profile, we change our coordinates again to $s=R-r$ (because the driving force is only non-zero inside the EDL) and use \eqref{eq:trick}, which makes it easier to integrate \eqref{eq:DifOsmStokes} twice and obtain
\begin{equation}
u_{\rm DO}(s)=-\frac{4k_{\rm B}T\lambda_D}{\eta}\partial_z\rho_s\left(\lambda_D\log\left(1-\gamma^2 e^{-2\kappa s}\right) + (2+c)s +d\right),
\end{equation}
where $c$ and $d$ are integration constants. Since all derivatives vanish on the channel axis ($s=R$), we have that the fluid flow must be constant outside of the EDL. This allows us to fix $c=-2$ such that the linear term cancels. The final constant $d$ can then be found by imposing the slip boundary condition for $u_z$. The solution to the diffusio-osmotic fluid flow is then found as
\begin{equation}
u(s)=-\frac{k_{\rm B}T}{2\pi \eta\lambda_B}\frac{\partial_z\rho_s}{\rho_s}\left(\log\left(\dfrac{1-\gamma^2 e^{-2\kappa s}}{1-\gamma^2}\right) + \frac{b}{2\lambda_D}P_2\right).
\end{equation}
We can write the diffusio-osmotic flow outside of the EDL, $u_{\rm DO}$, as
\begin{equation}\label{eq:uDO}
u_{\rm DO}=-\frac{\partial_z\mu}{4\pi \eta\lambda_B}\left(P_3+\frac{b}{\lambda_D}P_2\right)
\end{equation}
This allows us to calculate the volumetric flow rate due to diffusio-osmosis,
\begin{gather}
\begin{aligned}
Q_{\rm DO}=&\pi R^2u_{\rm DO}-\frac{\partial_z\mu}{2\pi \eta\lambda_B}2\pi R\int\limits_0^R\dif s\log\left(1-\gamma^2e^{-2\kappa s}\right)\\
=&-\pi R^2\frac{\partial_z\mu}{4\pi\lambda_B\eta}\left(\frac{b}{\lambda_D}P_2+P_3-\frac{\lambda_D}{2R}P_4\right)
\end{aligned}
\raisetag{2cm}
\end{gather}
and thus we find the next Onsager coefficient
\begin{equation}
L_{31}=\frac{1}{\pi R^2}\frac{Q_{\rm DO}}{-\partial_z \mu}=\frac{1}{4\pi\lambda_B\eta}\left(\frac{b}{\lambda_D}P_2+P_3-\frac{\lambda_D}{2R}P_4\right).
\end{equation}
By comparing $L_{31}$ with \eqref{eq:L13} we have that $L_{31}=L_{13}$ as it should. 


\subsection{Calculation $L_{23}$}

The diffusio-osmotic $I_{\rm DO}$ consists of two contributions, from diffusion ($I_{\rm DO, dif}$) and from advection ($I_{\rm DO, adv}$). The novel contribution to $L_{23}$ mentioned in the main text originates from $I_{\rm DO, dif}$,
\begin{gather}\label{eq:IDOdif}
\begin{aligned}
I_{\rm DO, dif}&=-2\pi e\int\limits_0^R\dif r r \left(D_+\partial_z\rho_+-D_-\partial_z\rho_-\right)\\
&=-4\pi De\partial_z\rho_s\int\limits_0^R\dif r r \left(-\sinh\phi+\beta(\cosh\phi-1)+\beta\right)
\end{aligned}
\raisetag{1.3cm}
\end{gather}
The first expression can be calculated using charge neutrality of the EDL and the second term is the integral $P_2$ defined above (\eqref{eq:integrals}). We thus find
\begin{equation}\label{eq:IDOdif1}
I_{\rm DO, dif}=-2\pi R^2\dfrac{D e}{k_{\rm B}T}\partial_z\mu\left(-\frac{\sigma}{R}+\rho_s\beta(1+2\frac{\lambda_D}{R}P_2)\right).
\end{equation}
The advective contribution to the electric current is given by
\begin{equation}
I_{\rm DO, adv}=2\pi Re\int\limits^R_0\dif s\rho_e u=-2\pi R \epsilon\int\limits_0^R\dif s \partial_s^2\psi u(s),
\end{equation}
We have already calculated the $u(s)$ above. To solve for $I_{\rm DO, adv}$ it is best to rewrite this expression by partial integrating it twice,
\begin{equation}
I_{\rm DO, adv}=-2\pi R  \epsilon\left(-\frac{\sigma_e}{\epsilon}u_0+\int\limits_0^R\dif s(\psi-\psi_0)\partial_s^2u(s)\right),
\end{equation}
where we used that $\partial_s u (s=R)=0$. Now we can plug in \eqref{eq:DifOsmStokes} to eliminate the $\partial^2_s u$. This leaves the integral defined above as $P_5$, and we can write $I_{\rm DO, adv}$ as
\begin{equation}\label{eq:IDOadv}
\frac{I_{\rm DO, adv}}{\pi R^2}=\frac{e\partial_z\mu}{2\pi\lambda_B\lambda_D\eta R}\left(b\sigma P_2+\dfrac{z_{\sigma}}{4\pi\lambda_B}P_5\right).
\end{equation} 
Now we can write down the next Onsager coefficient,
\begin{equation}
\begin{aligned}
L_{32}=&\frac{1}{\pi R^2}\frac{I_{\rm DO}}{-\partial_z \mu}=-\dfrac{2De}{k_{\rm B}T}\left(\frac{\sigma}{R}-\rho_s\beta\left(1+2\frac{\lambda_D}{R}P_2\right)\right)\\
&-\frac{e}{2\pi\lambda_B\lambda_D\eta R}\left(b\sigma P_2+\dfrac{1}{4\pi\lambda_B}P_5\right).
\end{aligned}
\end{equation}
Comparing $L_{32}$ with $L_{23}$ we see that the two coefficients are indeed equal, as required.

There is, however, a subtlety involved with the above computation. This problem becomes apparent if we write the first term of $I_{\rm DO,dif}$ differently and applying charge neutrality \eqref{eq:neutr} again,
\begin{equation}\label{eq:IDOdif2}
I_{\rm DO, dif}=-4\pi De\partial_z\left(\rho_s\int\limits_0^R\dif r r \left(-\sinh\phi+\beta(\cosh\phi-1)+\beta\right)\right),
\end{equation} 
Now, the problem only concerns the first term on the right hand side of \eqref{eq:IDOdif2}, so we omit the terms proportional to $\beta$ for clarity. We can namely use charge neutrality \eqref{eq:neutr} before calculating the derivative to obtain
\begin{equation}\label{eq:IDOdif3}
\begin{aligned}
I_{\rm DO, dif}&=-4\pi De\partial_z\left(\rho_s\int\limits_0^R\dif r r \left(-\sinh\phi\right)\right)\\
&=-\pi DeR^2\partial_z\left(-\frac{\sigma}{R}\right)=0,
\end{aligned}
\end{equation}
in the case of a constant $\sigma$. We thus find that, contrary to \eqref{eq:IDOdif1}, this term vanishes. Both cannot be correct, and there must be a faulty assumption underlying either \eqref{eq:IDOdif1} or \eqref{eq:IDOdif1}. \eqref{eq:IDOdif3} seems to be more exact, as it only relies on charge neutrality, which is probably the reason this has been adopted by previous studies \cite{Mouterde2018,Siria2013}. However, we have concluded in \eqref{eq:IDOdif3} that the derivative of this term vanishes even though the only $z$ dependence of comes from $\rho_s$, which is mathematically inconsistent. This does not imply that the charge neutrality condition is incorrect. On the contrary, in order for charge neutrality (\eqref{eq:neutr}) to be consistent we have that $\phi$ also depends on $z$ in such a way that \eqref{eq:neutr} will hold. \eqref{eq:IDOdif3} therefore only holds for a consistent analysis of diffusio-osmosis that incorporates the $z$ dependency of $\phi$. Interestingly, we find that we regain \eqref{eq:IDOdif1} from such an analysis, as we will show below. 

We have already shown that the fluid flow is unaffected by a laterally varying $\phi$, because the resulting lateral electric field $E_z$ cancels the electric body force in the Stokes equation. However, $E_z$ does contribute to the electric current.. To continue, we assume that we can still use the same Poisson-Boltzmann equations for $\phi$, but that this solution is now also a function of $z$ via $\rho_s$ and thus $\lambda_D$. This allows us to determine $E_z$ from \eqref{eq:PB}, which can be written in terms of the normal derivative $\partial_s\phi$,
\begin{equation}
\partial_z\phi=\frac{1}{2}\partial_s\phi\left(s+\frac{\lambda_D}{\cosh\frac{1}{2}\phi_0}\right)\partial_z(\log\rho_s).
\end{equation}
Although this electric field will not influence the fluid flow, and thus $Q_{\rm DO}$ and $I_{\rm DO, adv}$, we do obtain a novel, conductive contribution to the generated electric current,
\begin{equation}
\begin{aligned}
I_{\rm DO,con}&=-2\pi R e\int\limits_0^R\dif s (D_+\rho_++D_-\rho_-)\partial_z\phi\\
&=-4\pi RD\rho_s e\int\limits_0^R\dif s (\cosh\phi-\beta\sinh\phi)\partial_z\phi.
\end{aligned}
\end{equation}
In order to determine $I_{\rm DO, con}$, we first solve the integral
\begin{equation}
\begin{aligned}
n_1&=\int\limits_0^R\dif s (\cosh\phi-\beta\sinh\phi)\partial_s\phi\\
&=\int\limits_{\phi_0}^0\dif \phi (\cosh\phi-\beta\sinh\phi)\\
&=-\sinh\phi_0+\beta(\cosh\phi_0-1).
\end{aligned}
\end{equation}
The solution to this integral aids in solving the second integral
\begin{equation}
\begin{aligned}
n_2&=\int\limits_0^R\dif s (\cosh\phi-\beta\sinh\phi)s\partial_s\phi\\
&=\left[s\sinh\phi-s\beta\cosh\phi\right]^{s=R}_{s=0}-\int\limits_0^R\dif s(\sinh\phi-\beta\cosh\phi)\\
&=-\int\limits_0^R\dif s(\sinh\phi-\beta(\cosh\phi-1))=\frac{\sigma}{2\rho_s}+\beta\lambda_DP_2,
\end{aligned}
\end{equation}
where, in the last line, we have inserted the solutions to the integral $P_2$. The conductive contribution to the diffusio-osmotic current can, after some algebra, be written as
\begin{equation}
\begin{aligned}
I_{\rm DO,con}&=-2\pi R\partial_z\mu \frac{De}{k_{\rm B}T}\rho_s\lambda_D\left(\frac{n_1}{\cosh\frac{1}{2}\phi_0}+\frac{n_2}{\lambda_D}\right)\\
&=2\pi R^2\frac{De}{k_{\rm B}T}\partial_z\mu\left(\frac{\sigma}{R}-\frac{\sigma^{\ast}}{2R}\beta\left(P_2+\frac{\cosh\frac{1}{2}\phi_0-1}{\cosh\frac{1}{2}\phi_0}\right)\right).
\end{aligned}
\end{equation}
As discussed, the diffusive contribution to $I_{\rm DO, dif}$ must be calculated differently if the surface potential depends on $z$ too. Starting with \eqref{eq:IDOdif2} we find
\begin{equation}
\begin{aligned}
I_{\rm DO, dif}&=-4\pi De\partial_z\left(-R\sigma+\beta\left(R\rho_s\lambda_DP_2+\frac{1}{2}R^2\rho_s\right)\right)\\
&=-2\pi R^2 De\beta\left(\partial_z\rho_s+\frac{\sigma^{\ast}}{2R}\frac{\cosh\frac{1}{2}\phi_0-1}{\cosh\frac{1}{2}\phi_0}\partial_z\log\rho_s\right),
\end{aligned}
\end{equation} 
where we used that $\partial_z(\rho_s\lambda_DP_2)=\frac{\sigma^{\ast}}{4}\left(\frac{\cosh\frac{1}{2}\phi_0-1}{\cosh\frac{1}{2}\phi_0}\right)$ and $\partial_z\sigma=0$. Adding $I_{\rm DO,con}$ and $I_{\rm DO,dif}$ we find
\begin{equation}
I_{\rm DO, dif}+I_{\rm DO, con}=2\pi R^2\frac{De}{k_{\rm B}T}\partial_z\mu\left(\frac{\sigma}{R}-\beta\left(1+\frac{2\lambda_D}{R}P_2\right)\right),
\end{equation}
which is, interestingly, exactly the same as the expression we found using before, \eqref{eq:IDOdif}. Therefore we find that even though the surface will develop a lateral electric field due to the laterally varying EDL potential, this will not alter the final equations for $Q_{\rm DO}$ and $I_{\rm DO}$ we would get if we assume a constant surface potential. This means that we can safely ignore the $z$ dependence of $\phi$, and treat it as if it is a function of $r$ only. This gives the same result as if we would take this into account, but is much less laborious.

\subsection{Calculating $L_{33}$}

Lastly, we determine the diffusio-osmosic salt flux $J_{\rm DO}$. The salt flux has two contributions, one from diffusion and one from advection,
\begin{equation}
J_{\rm DO, dif}=-4\pi\dfrac{D}{k_{\rm B}T}\partial_z\mu\int\limits_0^R\dif r r\left(\cosh\phi-\beta\sinh\phi\right).
\end{equation}
These integrals are already discussed above, \eqref{eq:IEOcond}, so this allows us to write the diffusive contributions as
\begin{equation}
J_{\rm DO, dif}=-2\pi R^2\partial_z\mu\dfrac{D}{k_{\rm B}T}\rho_s\left(1+2\frac{\lambda_D}{R}P_2-\beta\frac{\sigma}{\rho_s R}\right).
\end{equation}
This leaves us to determine the advective contribution to the salt flux,
\begin{equation}
\begin{aligned}
J_{\rm exc, DO, adv}&=2\pi\rho_s\int\limits_0^R\dif r r(\rho_++\rho_-)u-2\rho_sQ_{\rm DO}\\
&=2\pi\int\limits_0^R\dif r r(\rho_++\rho_--2\rho_s)u.
\end{aligned}
\end{equation}
To continue, it is convenient to split up the fluid velocity in $u_{\rm DO}$ and a contribution that is only non-zero inside the EDL. Then we obtain
\begin{equation}
\begin{aligned}
J_{\rm exc, DO, adv}=&-\frac{\rho_s\partial_z\mu}{\lambda_B\eta}\Bigg(\left(P_3+\frac{b}{\lambda_D}P_2\right)\int\limits_0^R\dif s (\cosh\phi-1)\\
&+2\int\limits_0^R\dif s (\cosh\phi-1)\log\left(1-\gamma^2 e^{-\kappa s}\right)\Bigg),
\end{aligned}
\end{equation}
where we used the expression for $u_{\rm DO}$, \eqref{eq:uDO}. Here we encounter a final Poisson-Boltzmann integral $P_6$, \eqref{eq:integrals}, such that we find
\begin{equation}
J_{\rm exc, DO, adv}=-R\frac{\partial_z\mu}{\lambda_B\eta}\rho_s\lambda_D\left(P_2\left(P_3+\frac{b}{\lambda_D}P_2\right)-2P_6\right).
\end{equation}
This gives the final Onsager coefficient
\begin{equation}
\begin{aligned}
L_{33}=\frac{1}{\pi R^2}\frac{J_{\rm exc, DO}}{-\partial_z\mu}=&\dfrac{2D}{k_{\rm B}T}\left(\rho_s+\frac{2\rho_s\lambda_D}{R}P_2-\beta\frac{\sigma}{R}\right)\\
&+\frac{\rho_s\lambda_D}{\pi\lambda_B\eta R}\left(2P_2-4P_3+\frac{b}{\lambda_D}P_2^2\right).
\end{aligned}
\end{equation}

\end{document}